\documentclass[aps,prstab,superscriptaddress,floatfix,twocolumn,amsfonts,amsmath,amssymb,amsart]{revtex4-1}
\usepackage{tikz-feynman}
\usepackage{comment}
\usepackage{graphicx} 
\usepackage{float}
\usepackage{dcolumn} 
\usepackage{bm} 
\usepackage[colorlinks, breaklinks=true,linkcolor=red, citecolor=blue, linktocpage=true]{hyperref}
\usepackage{subcaption}
\renewcommand{\vec}{\mathbf}

\begin{document}

\title{Screening of the band gap in electrically biased bilayer graphene: From Hartree to Hartree-Fock}
\author{Jack N. Engdahl}
\affiliation{School of Physics, University of New South Wales, Sydney 2052, Australia}
\author{Zeb E. Krix}
\affiliation{Department of Physics, University of Basel, CH-4056 Basel, Switzerland}
\author{Oleg P. Sushkov}
\affiliation{School of Physics, University of New South Wales, Sydney 2052, Australia}
\date{\today}

	\begin{abstract}
	  {It is well known that a direct band gap may be opened in bilayer graphene
            via the application of a perpendicular  electric field (bias).
The bias and the chemical potential are controlled by electrostatic gating where the top and bottom gate voltages are tuned separately.
The value of the band gap opened by the bias field is influenced by the self screening of the bilayer graphene.             
The Hartree contribution to the self screening is well known in literature, with Hartree screening significantly renormalizing the gap.
In the present work we derive the Fock contribution to the self screening and demonstrate that
it is equally important and in the low density regime even more important than the Hartree contribution. We calculate the Hartree-Fock screened band gap as a function of electron doping at zero temperature and also as a function of temperature at zero doping.}
  
\end{abstract}

\maketitle
\section{Introduction}
The advent of graphene \cite{novoselov_electric_2004} has opened an avenue into the study of exciting new physics and led to renewed interest in two dimensional (2D) materials overall. Bilayer graphene has attracted considerable attention due to its unique electronic properties \cite{mccann_electronic_2013}, including the ability to tune the band structure. While intrinsic Bernal stacked bilayer graphene has simple parabolic band structure with no band gap, it is well known that a gap may be opened by breaking the inversion symmetry of the system by applying a perpendicular bias electric field, with this electric field generated either by asymmetrical chemical doping of the graphene layers \cite{ohta_controlling_2006,castro_biased_2007,zhou_substrate-induced_2007,alattas_band_2018} or with electrostatic gating \cite{zhang_direct_2009,guinea_electronic_2006,oostinga_gate-induced_2008,anderson_exploring_2023}. Such a system is referred to as biased bilayer graphene. 

The electronic properties of bilayer graphene, biased or not, render it an excellent platform for the study of many different physical phenomena including exciton physics \cite{yang_excitons_2011,ju_tunable_2017,li_excitonic_2017,henriques_absorption_2022,sauer_exciton_2022,liu_interference_2023,saleem_theory_2023,scammell_dynamical_2023,scammell_exciton_2023}, electron hydrodynamics \cite{gusev_viscous_2021,monch_ratchet_2022,cruise_observability_2024}, plasmonics \cite{grigorenko_graphene_2012,liu_interference_2023,fei_tunneling_2015} and superconductivity, both conventional \cite{anirban_superconductivity_2021,li_tunable_2024,wagner_superconductivity_2024,sboychakov_triplet_2023,zhou_isospin_2022} and unconventional \cite{cao_unconventional_2018,sharma_superconductivity_2020,cao_nematicity_2021,khalaf_charged_2021,chew_higher-order_2023,stepanov_competing_2021}. Bilayer graphene also lends itself to the field of quantum computing, where it is seen as an ideal material for the creation of quantum dots \cite{banszerus_electronhole_2020,banszerus_single-electron_2020,ge_visualization_2020,banszerus_spin-valley_2021,solomon_valley_2021,anderson_exploring_2023,hecker_coherent_2023,korkusinski_spontaneous_2023,saleem_theory_2023,garreis_long-lived_2024}. Regardless of the application it is imperative to accurately know the band structure including the band gap in the case of biased bilayer graphene and its response to changes in charge carrier density. For example, in the case of studying exciton physics in biased bilayer graphene the exciton binding energy is dictated by the band structure and the formation of an excitonic condensate is dependent on the relative size of the binding energy compared to the band gap. In this scenario it is clear that rescaling the band gap can greatly impact the physical phenomena observed in the system. 
 
In biased bilayer graphene the bilayer is located between two metallic gates with dielectrics in between the bilayer and the gates. The band gap is opened by a voltage asymmetry between the
top gate and bottom gate voltages.
Naively one would expect the gap to be equal to the bias induced potential difference between the two layers of graphene. However, charges within the bilayer screen the external bias field and as a result
the gap is reduced by a factor of around $2-3$ times. This screening factor depends on the
chemical potential (addition of conduction electrons). This is the Hartree screening
considered in Refs. \cite{mccann_asymmetry_2006, fogler_comment_2010,mccann_electronic_2013}.

There is an additional effect related to many body screening that reduces the band gap.
This effect is known in monolayer transition metal dichalcogenides (TMDs) where it has been observed 
experimentally as a significant reduction of the gap in a metallic TMD compared to that
in insulating TMD \cite{chernikov_population_2015,pogna_photo-induced_2016,cunningham_photoinduced_2017,yao_optically_2017,qiu_giant_2019,bera_atomlike_2021,kang_universal_2017,nguyen_visualizing_2019,liu_direct_2019}. 
This shrinking of the band gap has been explained theoretically as a consequence of renormalization of the Fock self energy \cite{gao_dynamical_2016,gao_renormalization_2017,liang_carrier_2015,faridi_quasiparticle_2021,engdahl_theory_2025}.
The Fock effect is technically complex as it is related to the RPA summation of an
infinite chain of Feynman diagrams. In our recent work  we
have developed an efficient imaginary frequency tecnique for evaluation of this
effect \cite{engdahl_theory_2025}.

In the current work we apply our method
to biased bilayer graphene. Both Hartree and Fock screening contribute to the value of
the band gap at  fixed gate potentials. Specifically we consider the following situations.\\
(i) Insulating bilayer graphene at zero temperature.  In this case the result is the value
of the gap at a fixed bias.\\
(ii) The dependence of the gap at zero temperature on the density of charge carriers
in the conduction (valence)  band. The Fock contribution
changes the dependence qualitatively, and quantitatively it results in sharp reduction
of the gap at low density with slower monotonic increase at higher density.\\
(iii) The temperature dependence of the gap in insulating graphene. The Fock contribution
changes the dependence qualitatively, and quantitatively it results in a sharp step-like reduction
of the gap at low temperature with slower monotonic increase at higher temperature.

The structure of the paper is as follows. In Sec.\ref{sec:setup} we describe the biased bilayer graphene system that we consider in this work and introduce the Hartree and Fock contributions to band gap renormalization. In Sec.\ref{sec:Hartree} we consider the band gap at zero temeprature. We first present a derivation of the Hartree
screening of the band gap, following Ref.\cite{fogler_comment_2010}.
We need this derivation for coherent explanation of the full Hartree-Fock
effect. We then discuss the polarization operator of bilayer graphene in Sec.\ref{sec:pol} and then derive the Fock contribution to band gap renormalization using the methods of QED in Sec.\ref{sec:Fock}. Following this in Sec.\ref{sec:results_n} we present plots of the Hartree-Fock screened band gap as a function of doping and show that the Fock correction is significant. In Sec.\ref{sec:BGR_T} we present a method for a simple experiment that can detect the effect outlined in the previous section. In this experiment one would begin with zero temperature insulating biased bilayer graphene and measure the resistivity as the temperature is increased. Rather than introducing mobile charge carriers into the system via electrostatic gating they are moved from the valence to conduction band via finite temperature, however the principle of band gap reduction is the same. We present theory altered for the special case of the finite temperature insulator and then show plots of the Hartree-Fock screened band gap vs temperature. We present our conclusions in Sec.\ref{sec:Con}.
 
\section{Band gap screening, Hartree and Fock contributions}
\label{sec:setup}
\begin{figure} [h]
          \vspace{-5pt}
 	\includegraphics[width=0.7\linewidth]{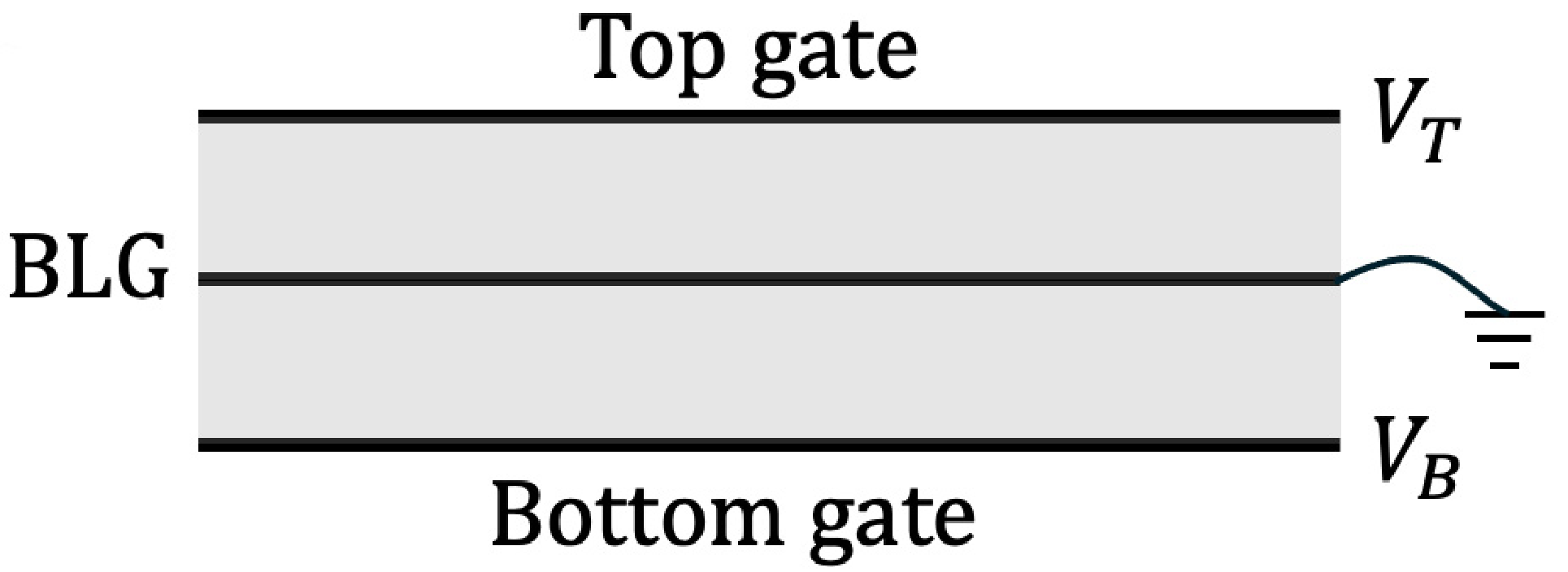}
        \vspace{10pt}
 	\caption{Schematic of the biased bilayer graphene setup with top and bottom gates
          at voltages $V_T$ and $V_B$. The bilayer graphene (BLG) is grounded.}
 	\label{fig:setup}
 \end{figure}

\begin{figure} [h]
 	\includegraphics[width=0.9\linewidth]{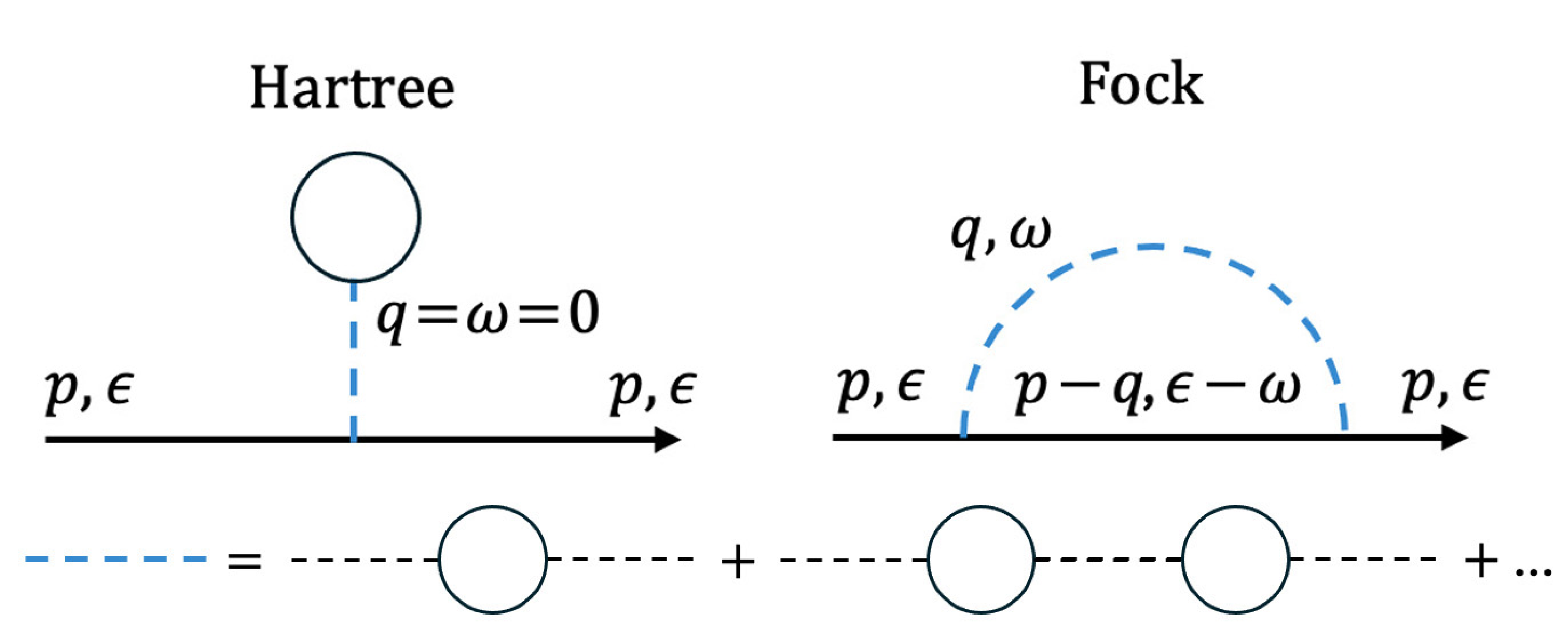}
        \vspace{10pt}
 	\caption{Hartree and Fock diagrams for BLG electron self energy $\Sigma$.
          The blue dashed line represents the dressed Coulomb interaction
          and the black dashed line is the bare Coulomb interaction.
          The circle is the electron polarization operator.
          }
 	\label{fig:diagr}
 \end{figure}

A schematic of the biased bilayer graphene setup we consider is presented in
Fig.~\ref{fig:setup}.
  The bilayer graphene (BLG) is grounded and both the electric bias and the
 chemical potential are controlled by voltages $V_T$ and $V_B$ at the top
 and bottom gates.
 The screening effect is described by the BLG electron self energy $\Sigma$
 given by diagrams shown in Fig.~\ref{fig:diagr}.
We will denote the band gap by ${\cal D}$ that is related to $\Delta$ that we introduce below as
\begin{eqnarray}
  \Delta=\frac{{\cal D}}{2}
\end{eqnarray}
Note that ${\cal D}$ and $\Delta$ can be positive or negative.
The Hartree contribution contains the Coulomb quantum at zero momentum, $q=0$.
Therefore, working in momentum space one needs to perform calculations at
$q\ne 0$ and then take the limit $q\to 0$. Therefore it is much more
convenient to work in coordinate space as it has been done in Refs.
\cite{mccann_asymmetry_2006, fogler_comment_2010,mccann_electronic_2013}.
On the other hand the Fock contribution is more convenient to evaluate in momentum space.

\section{Hartree contribution to the screening of the Band Gap}
\label{sec:Hartree}
\begin{figure} [h]
 	\centering
 	\includegraphics[width=0.8\linewidth]{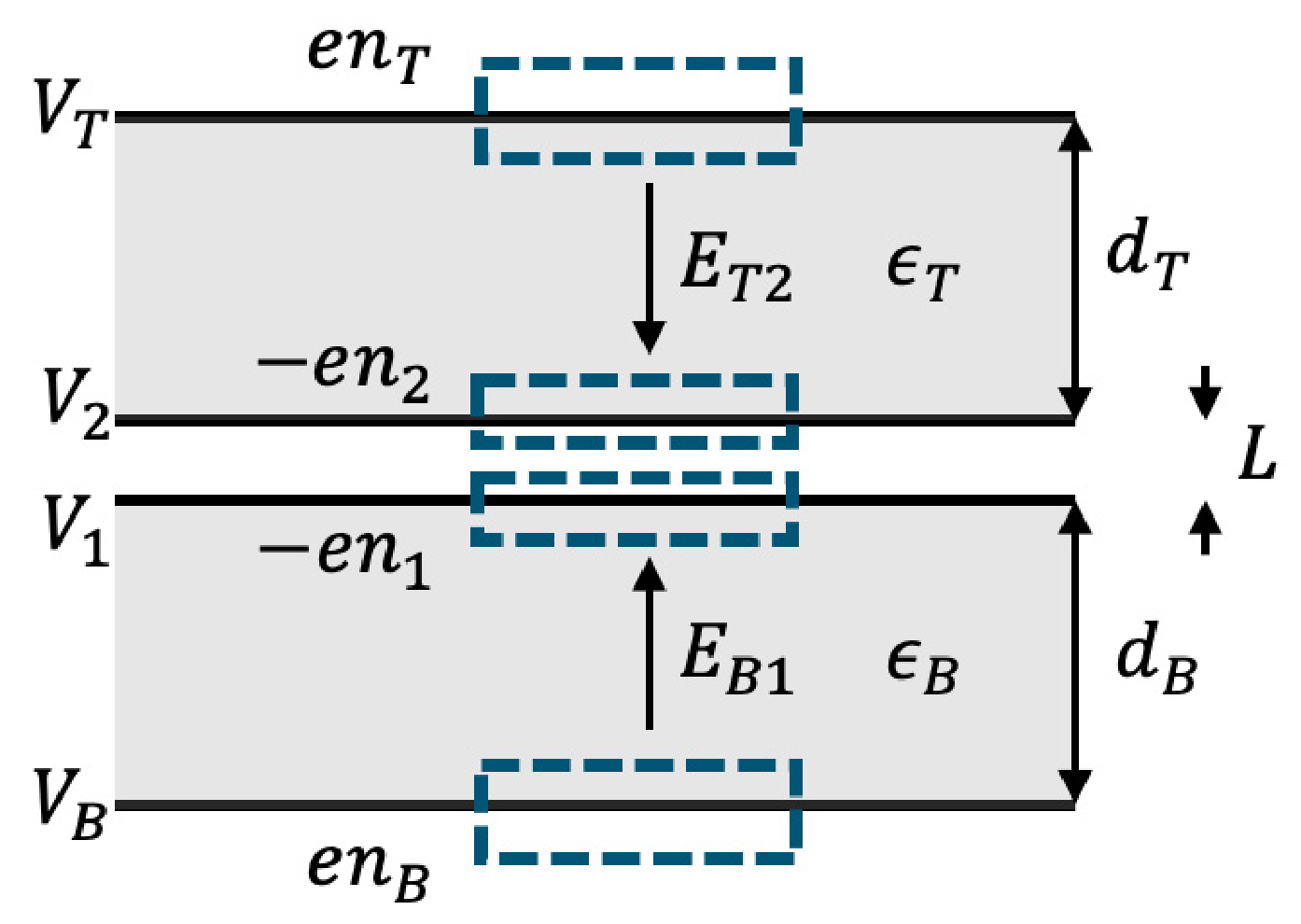}
 	\caption{Gaussian surfaces shown by dashed blue lines.}
 	\label{fig:gauss}
 \end{figure}

 Here we repeat calculations
 of Refs.\cite{mccann_asymmetry_2006, fogler_comment_2010,mccann_electronic_2013}.
 First we consider electrostatics of the biased BLG system. Fig.~\ref{fig:gauss} shows a more detailed
 picture of the setup. The BLG consists of two
 layers separated by distance $L=0.335$ nm \cite{mccann_electronic_2013} with electrostatic potentials
 $V_1$ and $V_2$.
 BLG is grounded which implies that the chemical potential is zero, $\mu=0$,
 however, $V_1$ and $V_2$ are not simultaneously zero. $E_{12}$ is the electric field between layers
 of the BLG. The elementary charge is $e=|e|$.
 We choose positive directions of electric fields $E_B$ and $E_T$ as it is
 shown in Fig.~\ref{fig:gauss}, hence $E_T=(V_T-V_2)/d_T$ and $E_B=(V_B-V_1)/d_B$,
 where $d_T$ is the  distance from the top gate to the graphene layer 2 and
$d_B$ is  the distance from the bottom gate to the graphene layer 1.
 Gate induced  number densites of electrons at BLG layers are $n_1,n_2$ and
 voltages related number densities of positive charges at the gates are
 $n_T,n_B$.
The dielectric constant between the bottom gate and the
 bilayer is $\epsilon_B$ and between the top gate and the bilayer is
 $\epsilon_T$. 
We have a condition of electroneutrality and Gauss' s law (in CGS units) for 
surfaces shown in Fig.~\ref{fig:gauss} by dashed lines.
\begin{eqnarray}
  \label{elst}
    &&n_B-n_1-n_2+n_T=0\nonumber\\
    &&\epsilon_BE_{B}=4\pi en_B\nonumber\\
    &&\epsilon_TE_{T}=4\pi en_T\nonumber\\
    &&-\epsilon_BE_{B} + E_{12} = -4\pi en_1\nonumber\\
    &&- E_{12} -\epsilon_TE_{T2} = -4\pi en_2.
     \end{eqnarray}
The self consistent bilayer gap is the electrostatic energy difference
between the BLG layers
\begin{equation}
 	{\cal D}=e(V_1-V_2)=eE_{12}L
\end{equation}
Here $L= 0.335$ nm is the interlayer distance \cite{mccann_electronic_2013}.
From Eqs.(\ref{elst}) one immediately gets
\begin{eqnarray}
  \label{nD}
   &&n = n_1+n_2 =\frac{\epsilon_B (V_B-V_1)}{4\pi ed_B} + \frac{\epsilon_T (V_T-V_2)}{4\pi ed_T}\\
&&{\cal D} = 2\pi e^2L(n_2-n_1) + \frac{eL}{2}\left[ \frac{\epsilon_B(V_B-V_1)}{d_B} - \frac{\epsilon_T (V_T-V_2)}{d_T} \right]\nonumber
 \end{eqnarray}
 For a realistic device $L\ll d_B,d_T$, hence $V_1,V_2\ll V_B,V_T$ (remember that
 BLG is grounded). Therefore Eqs.(\ref{nD}) can be rewritten as
\begin{eqnarray}
  \label{eq:Hartree_1}
  &&n = n_1+n_2 \approx\frac{\epsilon_B V_B}{4\pi ed_B} +
  \frac{\epsilon_T V_T}{4\pi ed_T}\nonumber \\
&&{\cal D} = 2\pi e^2L(n_2-n_1)+{\cal D}_{ext}\nonumber\\
&&{\cal D}_{ext} =  \frac{eL}{2}\left[ \frac{\epsilon_BV_B}{d_B} - \frac{\epsilon_T V_T}{d_T} \right]
 \end{eqnarray}
Here the first equation gives the total induced density of electrons in BLG.
${\cal D}_{ext}$ is the naive perturbation theory value of the gap,
and the first term in the middle equation is the Hartree screening
corresponding to the Hartree diagram in Fig.~\ref{fig:diagr}.

To determine $n_{1(2)}$ consider the low energy Hamiltonian of BLG  in the layer
representation
\begin{align}
  \label{hh}
 	\hat{H}
 	= \begin{bmatrix}
 		\Delta &
 		-\frac{1}{2m}\vec{p^2_-}\\
 		-\frac{1}{2m}\vec{p^2_+}&
 		-\Delta
 	\end{bmatrix}
\end{align}
Here $m\approx 0.033m_e$ is the effective mass \cite{zou_effective_2011}.
We remind that $\Delta={\cal D}/2$.
  Here $\vec{p_{+(-)}}=\tau p_x + ip_y$, where $\tau$ enumerates valley. The eigenenergies and normalized eigenfunctions of this Hamiltonian are
   \begin{eqnarray}
&& Valence \ band \ (VB): \ \ \ \ \epsilon=-\sqrt{\Delta^2+\frac{p^4}{4m^2}}\nonumber\\
&& Conduction \ band \ (CB): \ \ \epsilon=+\sqrt{\Delta^2+\frac{p^4}{4m^2}}
 \end{eqnarray}
 \begin{align}
 	\psi
 	= \sqrt{\frac{\epsilon+\Delta}{2\epsilon}} \begin{pmatrix}
 		1\\
 		-\frac{p^2e^{2i\tau\varphi}}{2m(\epsilon+\Delta)}
 	\end{pmatrix}
 	e^{i\vec{p}\cdot\vec{r}},
 \end{align}
 where $\varphi = \tan^{-1}(p_y/p_x)$. Noting this wavefunction is written in the basis of the layers $(2,\:1)^T$ corresponding to $($high energy, low energy$)^T$, we perform the usual summation of occupied states to find $n_{1(2)}$, where we account for $2\times2$ degeneracy from spin and valley degrees of freedom.
  \begin{align}
 	n_{1(2)}&=2\times2\times\frac{1}{2\pi}\int |\psi_{1(2)}|^2 pdp \nonumber\\
 	&=\frac{2}{\pi}\int\frac{\epsilon\mp\Delta}{2\epsilon}pdp
 \end{align} 
Here we assume that temperature is zero.
   Consider the partially filled conduction band and the fully filled valence band separately. For the conduction band we have
   \begin{align}
     \label{ncb}
 	n^{(CB)}_{1(2)}&=\frac{1}{\pi}\int^{p_F}_0 pdp \mp \frac{\Delta}{\pi}\int^{p_F}_0 \frac{pdp}{\epsilon} \nonumber\\
 	&=\frac{n}{2} \mp \frac{m\Delta}{\pi}\ln\left(\frac{\pi|n|}{2m|\Delta|}+\sqrt{1+\left(\frac{\pi|n|}{2m\Delta} \right)^2} \right).
 \end{align}
 Here we have used the Luttinger theorem
 \begin{eqnarray}
   \label{Lt}
   n=n_1+n_2=\frac{p_F^2}{\pi}.
 \end{eqnarray}
 For the valence band the term $\frac{1}{\pi}\int pdp$ has to be thrown away
 since it is compensated by positive ions of BLG. Hence we left with
 $\mp\frac{\Delta}{\pi}\int \frac{p}{\epsilon}dp$ that is
 logarithmically divergent because
 of the quadratic in momentum approximation in the low energy Hamiltonian (\ref{hh}), where this approximation is valid for $\epsilon \ll \gamma_1 \approx 400$ meV.
 Therefore we introduce the momentum ultraviolet cut off
 $\Lambda\sim \sqrt{2m\gamma_1}=2mv$, where
 $v=\frac{\sqrt{3}}{2}\gamma_0a \approx 10^6$ ms$^{-1}$ is the Fermi-Dirac
   velocity of the parent monolayer graphene.
(For definition of parameters $\gamma_0$, $\gamma_1$, $a$ see Ref.\cite{mccann_electronic_2013}.)
Hereafter for numerical estimmates we shall take $\Lambda=2mv$.
The valence band integration gives
\begin{align}
  \label{nvb}
 	n^{(VB)}_{1(2)}&=\mp\frac{\Delta}{\pi}\int_0^{\Lambda}\frac{pdp}{\epsilon}\nonumber\\
 	&\approx\pm \frac{m\Delta}{\pi}\ln\left( \frac{\Lambda^2}{m|\Delta|} \right).
 \end{align} 
  The valence band contribution is calculated in the logarithmic approximation,
  $ \ln\left( \frac{\Lambda^2}{m|\Delta|} \right) \gg 1$. Substituting in all parameter values and taking a realistic half band gap $\Delta~5-50$ meV the value of this logarithm is approximately $2.5-5$, hence the approximation gives a reasonable, but not necessarily very accurate estimate. Nevertheless we shall continue in this vein.
  Combining the conduction and valence band terms we obtain the layer densities 
  \begin{align}
    n_{1(2)}&=\frac{n}{2}\nonumber\\&\mp \frac{m\Delta}{\pi} \ln \left(\frac{\pi |n|}{2\Lambda^2} +\sqrt{\left( \frac{\pi n}{2\Lambda^2} \right)^2+
      \left(\frac{m\Delta}{\Lambda^2} \right)^2}\right).
 \end{align}
  Substituting $n_{1(2)}$ into electrostatic Eq.~\eqref{eq:Hartree_1} and
  writing in terms of the full band gap ${\cal D}=2\Delta$ we obtain the
  following self consistent equation for the Hartree screened gap.
  \begin{align}
    \label{sc1}
  &  {\cal D} ={\cal D}_{ext}+\delta{\cal D}_H(n,{\cal D})\\
&\delta{\cal D}_H(n,{\cal D})=  {\cal D}(2 e^2m L)
        \ln \left(\frac{\pi |n|}{2\Lambda^2} +
        \sqrt{\left( \frac{\pi n}{2\Lambda^2} \right)^2+
          \left(\frac{m{\cal D}}{2\Lambda^2} \right)^2}\right)\nonumber    
  \end{align}
  Here $\delta{\cal D}_H(n,{\cal D})$ is the Hartree screening.
  The coefficient in this term is, $2 e^2m L\approx 0.40$, which is non negligible. 
  Therefore, taking into account the logarithmic enhancement, the Hartree
  screening is significant, ${\cal D}_{ext}/{\cal D}(n=0) \approx 2-2.5$. It should be noted that this Hartree screening is entirely independent of
  the distance to the metallic gate and the dielectric environment surrounding
  the bilayer. We stress that the Hartree correction is always negative, it reduces the band gap.

  \section{RPA Coulomb interation}
  \label{sec:pol}
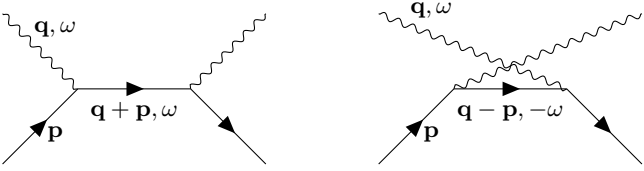
\begin{figure}[t]
 	\centering
 	\begin{tikzpicture} 
 		\begin{feynman}
 			\vertex at (1.75, 0.2)  {\(\vec{q}+\vec{p}, \omega\)};
 			\vertex at (0.7, 1.3)  {\(\vec{q}, \omega\)};
 			\vertex at (0.7, -0.1)  {\(\vec{p}\)};
 			\vertex at (1.0, 0.5) (a);
 			\vertex at (2.5, 0.5) (b);
 			\vertex at (0.0, -0.5) (c);
 			\vertex at (3.5, -0.5) (d);
 			\vertex at (0.0, 1.5) (e);
 			\vertex at (3.5, 1.5) (f);
 			
 			\vertex at (6.75, 0.2)  {\(\vec{q}-\vec{p}, -\omega\)};
 			\vertex at (5.7, 1.55)  {\(\vec{q}, \omega\)};
 			\vertex at (5.7, -0.1)  {\(\vec{p}\)};
 			\vertex at (6.0, 0.5) (g);
 			\vertex at (7.5, 0.5) (h);
 			\vertex at (5.0, -0.5) (i);
 			\vertex at (8.5, -0.5) (j);
 			\vertex at (5.0, 1.5) (k);
 			\vertex at (8.5, 1.5) (l);

 			\diagram*{
 				(c) -- [fermion] (a) -- [fermion] (b) -- [fermion] (d), (e) -- [boson] (a), (f) -- [boson] (b),
 				
 				(i) -- [fermion] (g) -- [fermion] (h) -- [fermion] (j), (k) -- [boson] (h), (l) -- [boson] (g)
 			};
 		\end{feynman}
 	\end{tikzpicture}
 	\vspace{0.2cm}
 	\caption{Diagrams for the Coulomb polarisation operator in BLG.  The straight lines correspond to electrons and the wiggly lines correspond to the Coulomb interaction.}
 	\label{fig:pol_open}
 \end{figure}

 For evaluation of the Fock diagram in Fig.~\ref{fig:diagr} one needs the full
 polarization operator as a function of momentum and frequency.
 In BLG the band gap is usually relatively small, on the order of
 ${\cal D} \sim 10-100$ meV. Therefore, unlike TMDs, both intra-band and
 inter-band virtual transitions contribute to the polarization.
 Hence, the polarization operator is significant even in the insulator.
The  polarization operator $\Pi(q,\omega)$ is given by the two digrams
shown in Fig.~\ref{fig:pol_open}. It is convenient to evaluate the polarization operator with these diagrams rather than the usual bubble diagram, although the final result is the same in both cases.
 
Here ${\bf q},\omega$ are momentum and frequency respectively
running through the Coulomb interaction, ${\bf p}$ is the momentum of the occupied
electron state and ${\bf p+q}$ is the momentum of the unoccupied
electron state. Hence, the polarization operator reads
\begin{align}\label{eq:pol_general}
 	\Pi^{\mu,\nu}(q,\omega) = 2\times2\int &\frac{d^2p}{(2\pi)^2} \bigg[ \frac{n_p^\mu(1-n_{p+q}^\nu)F^{\mu,\nu}_{p,p+q}}{\omega + \epsilon_p^\mu - \epsilon_{p+q}^\nu+i0} \nonumber\\&+ \frac{n_p^\mu(1-n_{p-q}^\nu)F^{\mu,\nu}_{p,p-q}}{-\omega + \epsilon_p^\mu - \epsilon_{p-q}^\nu + i0}\bigg] 
 \end{align}
Here $\mu,\nu$ correspond to the valence (-) or conduction (+) band,
for example $\Pi^{++}$ is the contribution to polarization that comes from
virtual transitions inside the conduction band, whereas
$\Pi^{-+}$ is the contribution that comes from
virtual transitions from the valence to the conduction band.
The common factor $2\times2$ accounts for  spin and valley degeneracy,
$n_p^\mu$ is the Fermi-Dirac  distribution, and
$F^{\mu\nu}_{p,p+q}=(\psi_{p+q}^\nu)^\dag\psi_p^\mu$ is the Coulomb vertex factor corresponding to wavefunction overlap.
Note that Eq. (\ref{eq:pol_general}) gives the Feynman polarization operator
that is reflected in $+i0$ in the denominators.
To be specific at zero temperature we consider only the insulator or the system
with electrons in conduction band. In this case only $\Pi^{-+},\Pi^{++}\ne 0$,
hence
  \begin{equation}
 	\Pi(q,\omega) = \Pi^{-+}(q,\omega) + \Pi^{++}(q,\omega).
 \end{equation}
In this case one can perform Wick's rotation to the imaginary frequency axis,
$\omega \to i\xi$. 
 \begin{align}\label{eq:pi}
  \Pi^{-+}(q,i\xi) &= 4\int \frac{d^2p}{(2\pi)^2} \frac{2(-\epsilon_p-\epsilon_{p+q})n_p^-(1-n_{p+q}^+)F^{-+}_{p,p+q}}{\xi^2+(-\epsilon_p-\epsilon_{p+q})^2}\nonumber\\
  \Pi^{++}(q,i\xi) &= 4\int \frac{d^2p}{(2\pi)^2} \frac{(\epsilon_p-\epsilon_{p+q})(n_p^+-n_{p+q}^+)F^{++}_{p,p+q}}{\xi^2+(\epsilon_p-\epsilon_{p+q})^2}.
  \nonumber\\
  \epsilon_p&=\sqrt{\Delta^2+\frac{p^4}{4m^2}}
   \end{align}
Here $\epsilon_p= \epsilon_p^+ = -\epsilon_p^-$.
Of course in the case of the zero temperature insulator $\Pi^{++}(q,i\xi)=0$ as there are no charge carriers in the conduction band, $n_p^+=0$.
 
The bare Coulomb interaction is $2\pi e^2/(\varepsilon q)$, where
$\varepsilon=(\epsilon_B+\epsilon_T)/2$ is the effective 2D dielectric constant.
RPA screening and metallic gate screening brings the interaction to the following form.
  \begin{eqnarray}
 	\label{V_q}
 	V_q(\omega)&=&
 	\frac{2\pi e^2}{\varepsilon q/\tanh(qd)- 2\pi e^2\Pi(q,\omega)+i0}
 \end{eqnarray}
  This corresponds to the blue dashed line in Fig.~\ref{fig:diagr}.
  Here $\tanh(qd)$ comes from the gate screening, which is easily derived using the method of images. For simplicity we have assumed $d_T=d_B=d$.

  \section{Electron self energy, Wick's rotation of the frequency integration
  contour} \label{sec:Fock}
 \begin{figure} [t]
 	\centering
 	\includegraphics[width=0.49\linewidth]{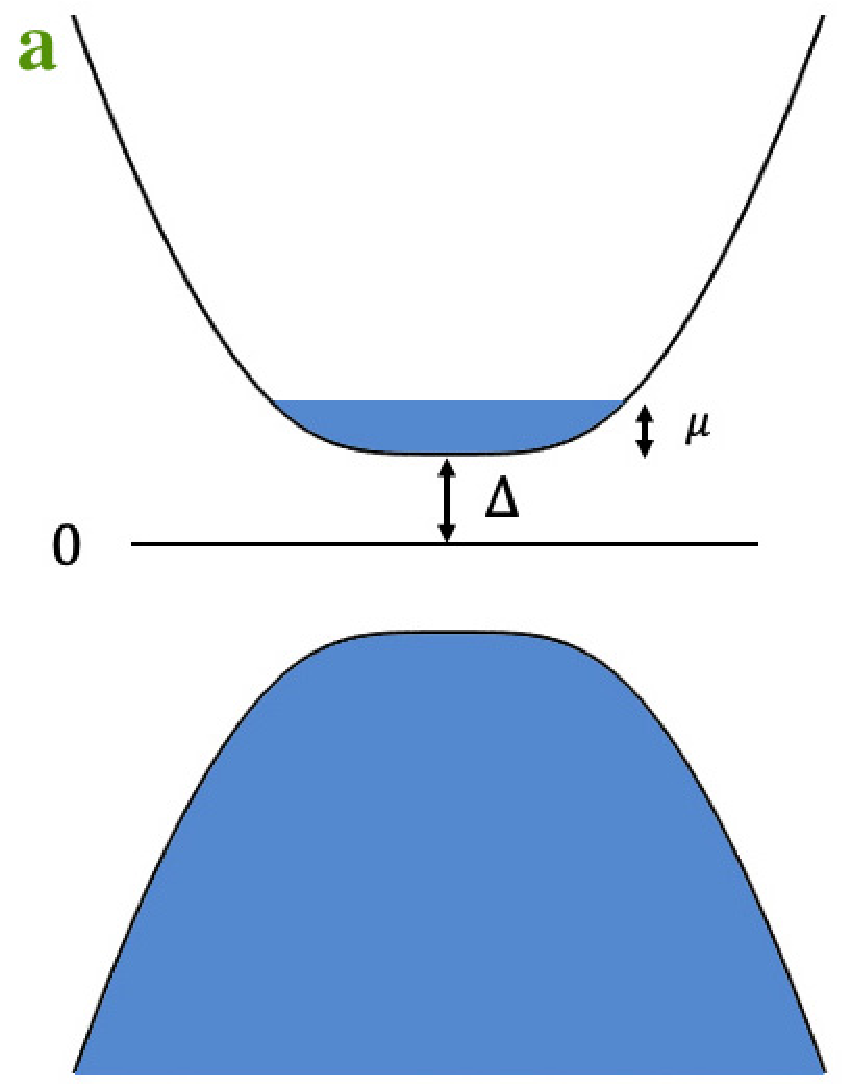}
 	\includegraphics[width=0.49\linewidth]{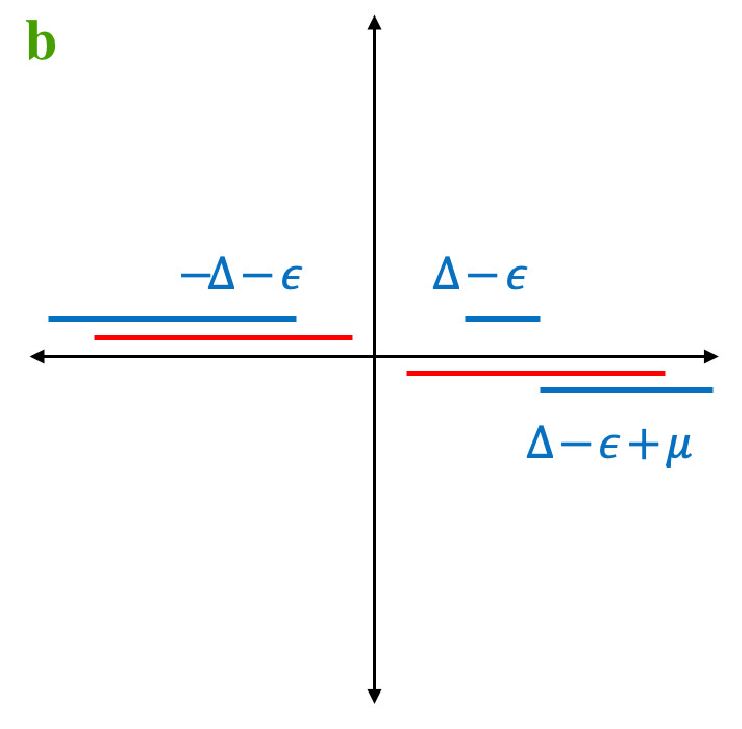}
 	\vspace{0.3cm}
 	\caption{Panel a: Band filling in metallic BLG with electrons in conduction band.   Panel b: cuts of the integrand Eq.~\eqref{se1} in the complex
          $\omega$ plane. Singularities of the Green's function
          $G^{(m)}(\epsilon+\omega)$ are shown in blue and singularities of
          the RPA interaction $V$ are shown in red.
}
 	\label{fig:diagram_m}
 \end{figure}
  In this section we consider the electron  self energy in the zero temperature metal.
  We follow a similar method to that which the authors developed for monolayer TMDs in  Ref.\cite{engdahl_theory_2025}.
 The physics behind the Fock correction to the band gap is exactly the same as that in monolayer TMDs, however the details of the calculation are somewhat different because of (i) a relatively small gap, (ii) a qualitatively different dispersion.

The self energy diagram is presented in Fig.~\ref{fig:diagr} Fock, where $\vec{p}$, $\epsilon$ are the external
momentum and frequency and $\vec{q}$, $\omega$ are internal momentum and
frequency.
  In the Feynman technique the self energy reads         
 \begin{align}
 	\label{se1}
 	\Sigma_p(\epsilon) = -\sum_{\bf q}\int G_{\vec{p}+\vec{q}}(\epsilon+\omega)
 	V_q(\omega)\frac{d\omega}{2\pi i}.
 \end{align}
 $G$ is the electron Green's function and $V_q(\omega)$ is
 the RPA screened Coulomb interaction (\ref{V_q}),
 (the blue  dashed line in Fig.~\ref{fig:diagr}).
 The Feynman electron Green's function for insulating BLG in the layer basis
 follows from the low energy effective Hamiltonian
(\ref{hh})
 \begin{eqnarray}
   \label{gf1}
   G_\vec{p}^{(ins)}(\epsilon) &=& \frac{\epsilon + \frac{p_+^2\sigma_+}{2m} +
     \frac{p_-^2\sigma_-}{2m} + \Delta \sigma_z}{\epsilon^2 - \epsilon_p^2 +i0}\nonumber\\
   &\to&
   \frac{\epsilon + \Delta \sigma_z}{2\epsilon_p} \left(\frac{1}{\epsilon - \epsilon_p +i0} -
   \frac{1}{\epsilon + \epsilon_p-i0}  \right)
    \end{eqnarray}
 To evaluate the band gap renormalization we need to calculate the self energy only at $p=0$.
 In this case $G_{\vec{p}+\vec{q}}$ in (\ref{se1}) becomes $G_{\vec{q}}$ and hence
 $q_\pm^2$ terms integrate to zero in the integrations over angles of $\vec{q}$.
 Therefore we omit $p_\pm^2$ terms in the electron Green's function, which leads to the second line in Eq.~(\ref{gf1}).
To be specific let us consider a metal with some concentration of electrons in the conduction band. For ease of notation for the remainder of this paper we replace the conduction electron density by $n_p^+\to n_p$.  The
 conduction band term in (\ref{gf1}) must be replaced by
 \begin{eqnarray}
   \label{repl1}
   \frac{1}{\epsilon - \epsilon_p +i0} \to \frac{n_p}{\epsilon - \epsilon_p -i0} + \frac{(1-n_p)}
        {\epsilon - \epsilon_p +i0}.
\end{eqnarray}
Hence, using the Sokhotski–Plemelj theorem $\frac{1}{x-i0} = \frac{1}{x} + \pi i\delta(x)$, the
Green's function for metallic BLG can be written as
\begin{align}
  \label{gf2}
   G^{(m)}_\vec{p}(\epsilon) = G^{(ins)}_\vec{p}(\epsilon) + \frac{\epsilon + \Delta \sigma_z}{2\epsilon_p}
   2\pi i n_p \delta(\epsilon -\epsilon_p)
 \end{align}
Note that $\frac{\epsilon + \Delta \sigma_z}{2\epsilon_p}$ in Eqs.(\ref{gf1}), (\ref{gf2}) in essence is the
projection operator; at $p=0$ and $\epsilon=\Delta$ it projects to the conduction band and at
$\epsilon=-\Delta$ it projects to the valence band.

With Eqs.~\eqref{se1}, \eqref{gf2} the  self energy reads
\begin{align}
  \label{sig1}
 	\Sigma_p(\epsilon) 
 	=&- \sum_{\bf q}\Bigg[\int G^{(ins)}_{\vec{p}+\vec{q}}(\epsilon+\omega)
 	V_q(\omega)\frac{d\omega}{2\pi i} \nonumber\\ 
 	&+ n_{p+q} \left( \frac{\epsilon_{p+q} + \Delta \sigma_z}{2\epsilon_{p+q}}\right) V_q(\epsilon_{p+q}-\epsilon) \Bigg]
 \end{align}

 Numerical frequency integration  in the self energy with real frequency $\omega$
 is mathematically inconvenient because the integrand has multiple singularities
 on the real axis. The usual way to overcome this problem is through rotation of the
 contour of integration to the imaginary axis.
 In Panel b of Fig.~\ref{fig:diagram_m} we show cuts of the
 integrand in the complex $\omega$-plane, for
 $-\Delta \leq \epsilon \leq \Delta$. Blue lines show cuts that originate
 from the electron Greens function $G(\epsilon+\omega)$ and the red lines
 show cuts that originate from the interaction $V(\omega)$.
 The branch cut in the first quadrant prevents straightforward application
 of Wick's rotation. This branch cut originates from conduction electrons
 in the Green's function, therefore it does not exist in the insulator where Wick's
 rotation is straightforward.
 The representation of the self energy shown in Eq.~\eqref{sig1} solves this problem
as one only needs to perform Wick's rotation on the first term in this equation. Rotation to the imaginary axis gives
 \begin{align}
   \label{sig2}
 	\Sigma_p(\epsilon) = &- \sum_{\bf q}\int G^{(ins)}_{\vec{p}+\vec{q}}(\epsilon+i\xi)
 	V_q(i\xi)\frac{d\xi}{2\pi} \nonumber\\ 
 	&-\sum_{\bf q} n_{p+q} \left( \frac{\epsilon_{p+q} + \Delta \sigma_z}{2\epsilon_{p+q}}\right)
 	V_q(\epsilon_{p+q}-\epsilon).
 \end{align}
 
  \begin{figure*} [ht!]
 	\centering
 	\includegraphics[width=0.41\linewidth]{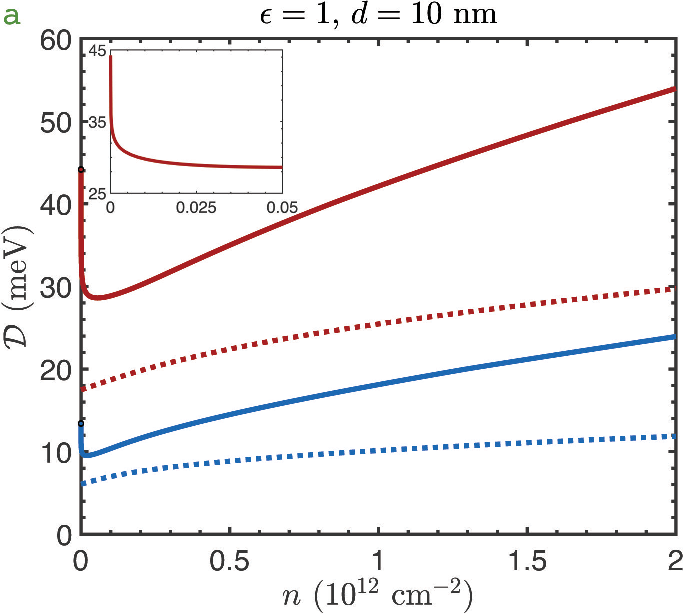}
 	\hspace{0.3cm}
 	\includegraphics[width=0.41\linewidth]{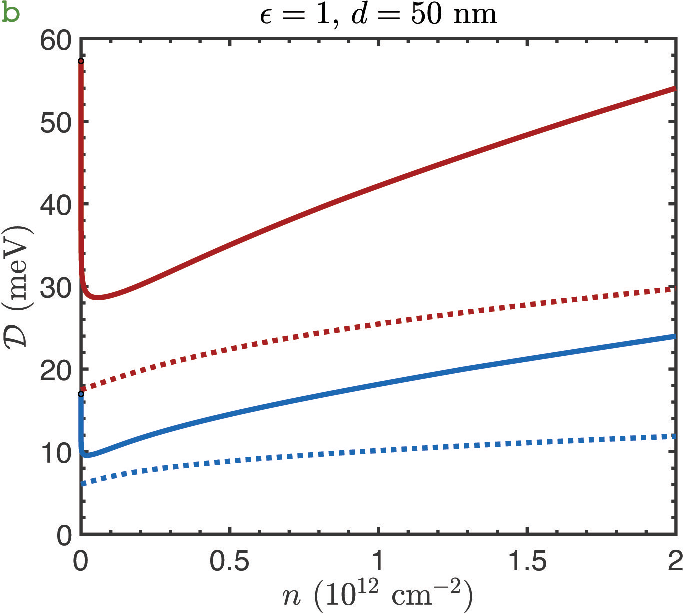}
 	\vspace{0.2cm}
 	\includegraphics[width=0.41\linewidth]{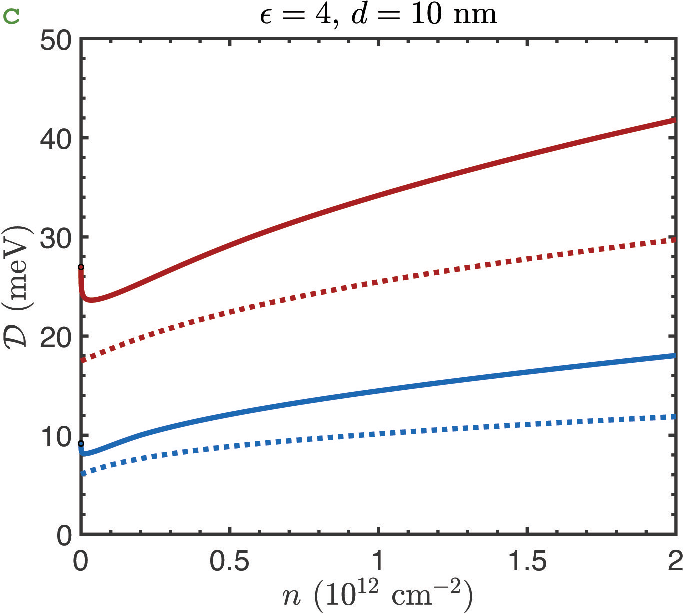}
 	\hspace{0.3cm}
 	\includegraphics[width=0.41\linewidth]{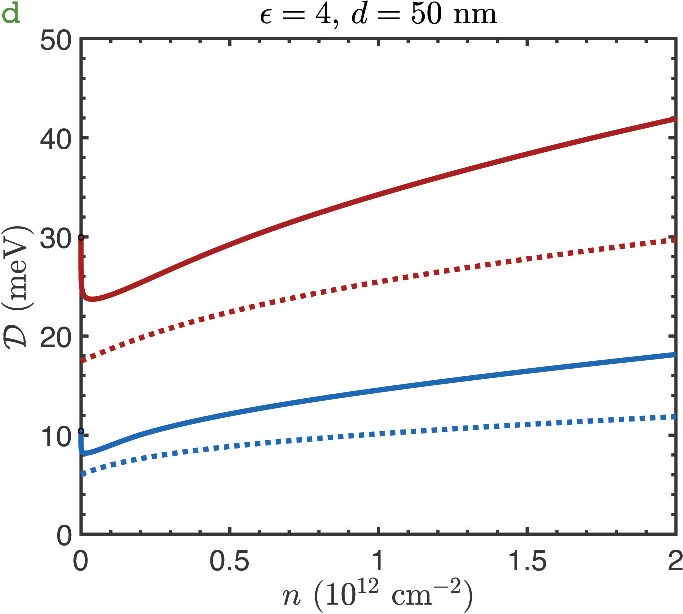}
 	\vspace{0.2cm}
 	\caption{Band gap of metallic biased bilayer graphene vs carrier density in the zero temperature approximation.
 		Panel a corresponds to $\epsilon =1$ and $d=10$ nm, the inset zooms in the very low density behavior $\propto n^{1/5}$.
 		Panel b corresponds to $\epsilon =1$ and $d=50$ nm. Panel c corresponds to $\epsilon =4$ and $d=10$ nm and panel d corresponds to $\epsilon =4$ and $d=50$ nm. Blue lines are for the external bias ${\cal D}_{ext}=20$ meV and red lines are for the external bias ${\cal D}_{ext}=50$ meV. Solid lines correspond to the Hartree-Fock screening  whereas dashed lines correspond to Hartree
 		screening only. The dots at the vertical axis at n=0 show the
 		gap value in the insulator, $n=0$.}
 	\label{fig:Delta_n}
 \end{figure*}
 
 The self energy is a $2\times 2$ matrix in the basis $(CB,\;VB)$. The Fock correction to the gap is
 \begin{eqnarray}
   \delta{\cal D}_F = 	\Sigma_{11}(p=0,\epsilon=\Delta) -
   \Sigma_{22}(p=0,\epsilon=-\Delta).
 \end{eqnarray}
 Note that both terms proportional to the identity matrix and terms proportional to $\sigma_z$ contribute to band gap renormalization. Following some algebraic manipulation, one arrives at
 \begin{align}
   \label{eq:delta_m}
 	\delta{\cal D}_F = &4\sum_q \int_0^\infty V_q(i\xi) \bigg[\frac{ \Delta(\epsilon_q^2 -\Delta^2)}{(\xi^2+\Delta^2+\epsilon_q^2)^2-4\Delta^2\epsilon_q^2}\bigg]\frac{d\xi}{2\pi} \nonumber\\
 	&-\sum_{\bf q} n_{q}\left( \frac{\epsilon_{q} + \Delta}{2\epsilon_{q}}\right)
 	V_q(\epsilon_{q}-\Delta).
 \end{align}
 Numerical evaluation of the integrals in this formula is straightforward
 as the integrands are smooth functions in $q$ and $\xi$. The first term in Eq.~\eqref{eq:delta_m} dominates the
 band gap correction. Similar to monolayer TMDs the Fock correction features a very sharp reduction of the
band gap at low electron density $n$. Using the low momentum limit for the polarization operator presented in Appendix.\ref{pi_approx}, we perform low density analysis of the integral Eq.~\eqref{eq:delta_m} in Appendix.\ref{low_density} and find the low doping $n$ power scaling
 \begin{eqnarray}
   \label{sharp}
   \delta{\cal D}_F(n=0) -\delta{\cal D}_F(n) \propto n^{1/5}.
   \end{eqnarray}
 This $1/5$ power scaling comes entirely from the first term in Eq.~\eqref{eq:delta_m}, the second term scales as $n^{2}$ and may be discarded in the regime of $n\to 0$.
 The density dependence \eqref{sharp} is even sharper than that in TMDs where
 $\delta{\cal D}_F(n=0) -\delta{\cal D}_F(n) \propto n^{1/3}$, see Ref.\cite{engdahl_theory_2025}. It is important to note that the sharp density dependence \eqref{sharp} is due to dynamical screening from mobile charge carriers. This is in contrast to the Hartree contribution which is only due to static screening. 

 \section{Band gap in zero  temperature insulator and zero temperature
   metal}\label{sec:results_n}
The band gap follows from numerical solution of the self consistent equation 
 \begin{equation}
   \label{sc2}
	{\cal D}(n) = {\cal D}_{ext} + \delta{\cal D}_H(n,{\cal D}) + \delta{\cal D}_F(n,{\cal D}).
 \end{equation}
 The density of conduction electrons $n$ and the ``external''
 gap ${\cal D}_{ext}$ (the gap which results from external bias in the absence of screening)  are determined by the gate volatages according to Eq.(\ref{eq:Hartree_1}). The Hartree screening $\delta{\cal D}_H(n,{\cal D})$ is given by Eq.(\ref{sc1}) and the Fock screening $\delta{\cal D}_F(n,{\cal D})$ is given by
 Eq.(\ref{eq:delta_m}).

 To present our results we fix the external bias ${\cal D}_{ext}$, the
 dielectric constant $\varepsilon$ and the distance to the gates $d$. We plot the
 physical gap as a function of conduction electron density.
 These plots are presented in Fig.~\ref{fig:Delta_n}, where solid lines
 include both Hartree and Fock screening while dashed lines show Hartree
 screening only. Red lines correspond to ${\cal D}_{ext}=50$ meV and blue lines correspond to ${\cal D}_{ext}=20$ meV. Curves are shown for suspended biased bilayer graphene ($\varepsilon=1$) and biased bilayer graphene
 encased in hBN ($\varepsilon=4$) with a metallic gate located at distance
 $d=10,\;50$ nm. The plots demonstrate that  the Fock screening is quantitatively very
 important, changing the gap compared to its Hartree value by a factor $\sim 2$.
 Moreover, the Fock screening gives a qualitative effect, with a very sharp reduction
 of the gap at low conduction electron density, see Eq.~\eqref{sharp}.
 \begin{figure*} [t!]
	\centering
	\includegraphics[width=0.45\linewidth]{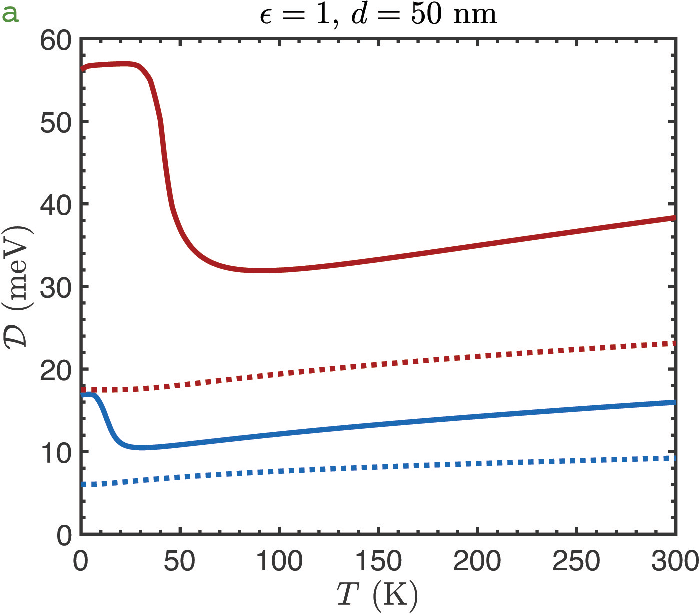}
	\hspace{0.1cm}
	\includegraphics[width=0.45\linewidth]{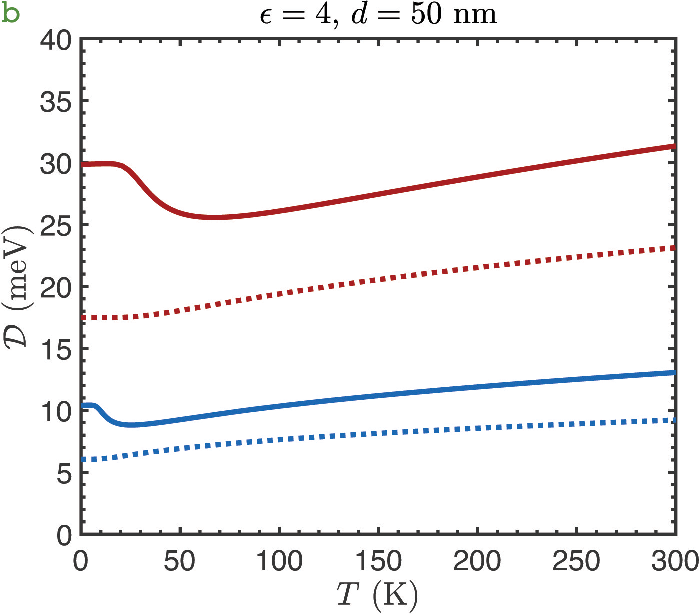}
	\caption{Band gap of insulating biased bilayer graphene vs temperature for biased bilayer graphene. Panel a corresponds to $\epsilon =1$ and $d=50$ nm and panel b corresponds to $\epsilon =4$ and $d=50$ nm. Blue lines are for ${\cal D}_{ext}=20$ meV and red lines are for ${\cal D}_{ext}=50$ meV. Solid lines correspond to the Hartree-Fock model of screening whereas dashed lines correspond to Hartree screening only.} 
	\label{fig:Delta_T}
\end{figure*}
  \section{Temperature dependence of the band gap in the insulator
} \label{sec:BGR_T}
The very sharp density dependence at low $n$ demonstrated in the previous section is due to the enormous polarizability of the BLG electron gas at low density. This hints that finite temperature might also influence the gap significantly,
even at a low density of thermally excited electrons. This effect
might be observed in transport experiments. To investigate this idea in the present
section we consider insulating BLG at a finite temperature.
The system remains electrically neutral, therefore the chemical potential
is zero, $\mu=0$.

 \subsection{Hartree term for finite temperature}
 At a finite temerature and $\mu=0$ Eqs. (\ref{ncb}), (\ref{nvb}) become
   \begin{align}
 &    n^{(CB)}_{1(2)}(T)=\frac{1}{\pi}\int^{\infty}_0 n_p pdp \mp \frac{\Delta}{\pi}\int^{\infty}_0 n_p \frac{pdp}{\epsilon_p} \nonumber\\
&     n^{(VB)}_{1(2)}(T)=\frac{2}{\pi}\int\frac{\epsilon\mp\Delta}{2\epsilon}
     (1-n_p)pdp  \nonumber\\
     &\to -\frac{1}{\pi}\int^{\infty}_0 n_p pdp
     \pm \frac{m\Delta}{\pi}\ln\left( \frac{\Lambda^2}{m|\Delta|}\right)
     \mp \frac{\Delta}{\pi}\int^{\infty}_0 n_p \frac{pdp}{\epsilon_p}
     \nonumber\\
 	n_p&= \frac{1}{e^{\epsilon_p/k_BT}+1}    
 \end{align} 
 This gives
   \begin{eqnarray}
     n_2-n_1=-\frac{m\Delta}{\pi}\ln\left(\frac{\Lambda^2}{m\Delta}\right)
     +\frac{4\Delta}{\pi}\int_0^\infty n_p\frac{pdp}{\epsilon_p}
   \end{eqnarray}
   Substitution to the electrostatic Eq.(\ref{eq:Hartree_1}) gives the Hartree
   correction
   \begin{eqnarray}
     \label{dHT}
     \delta{\cal D}_H&=&-2e^2mL{\cal D}
     \ln\left(\frac{2\Lambda^2}{m{\cal D}}\right)
     +8e^2L\Delta\int_0^\infty n_p\frac{pdp}{\epsilon_p}\nonumber\\
     &\approx& -2e^2mL{\cal D}
     \left[\ln\left(\frac{2\Lambda^2}{m{\cal D}}\right)
     -4\sqrt{\frac{\pi T}{2 {\cal D}}}e^{-{\cal D}/(2T)}  \right]       
\end{eqnarray}
   Here the integral $\int_0^\infty n_p\frac{pdp}{\epsilon_p}$ has been
   calculated analytically in the low temperature limit.
   It can be calculated numerically exactly at arbitrary temperature, all plots shown are calculated in this way. Plots of the gap versus temperature with account of the Hartree correction
   at fixed external bias ${\cal D}_{ext}$ are presented in Fig.~\ref{fig:Delta_T} by dashed lines.
At constant bias field the Hartree screening term causes the band gap to increase as temperature is increased. As is the case with the zero temperature metal, the Hartree contribution to the band gap variation is independent of substrate dielectric constant and metallic gate distance.

\subsection{Fock term for finite temperature} 
There are two reasons why temperature influences the Fock screening:
(i) The frequency integration must be replaced by Matsubara summation,
(ii) Electrons thermally excited from valence to the conduction band
have enormous polarizability and therefore influence the polarization operator.
The first reason is insignificant. Replacement of integration by summation does not significantly influence our results. therefore, we only consider the second reason in this subsection.

Consider the general expression for the polarization operator, Eq.~\eqref{eq:pol_general}. In the case of finite temperature all transitions $-+$, $--$, $+-$ and $++$ are allowed.
\begin{equation}
  \label{ehp}
	\Pi(q,i\xi) = \Pi^{-+}(q,i\xi) + \Pi^{++}(q,i\xi) + \Pi^{+-}(q,i\xi) + \Pi^{--}(q,i\xi)
\end{equation}
 The dominating contribution to the temperature effect comes from
$ \Pi^{++}(q,i\xi) = \Pi^{--}(q,i\xi)$.
 The expression for the Fock correction to the band gap immediatelly follows
 from  Eq.~\eqref{eq:delta_m}, where the polarization is replaced by Eq.~\eqref{ehp} and the second term is omitted as we consider the finite temperature insulator.
\begin{align}\label{eq:delta_T}
	\delta{\cal D}_F(T)&=4\sum_q \int_0^\infty V^{(T)}_q(i\xi) \nonumber\\
	&\times \bigg[\frac{ \Delta(\epsilon_q^2 -\Delta^2)}{(\xi^2+\Delta^2+\epsilon_q^2)^2-4\Delta^2\epsilon_q^2}\bigg]\frac{d\xi}{2\pi} \nonumber\\
\end{align}

\subsection{Temperature dependence of the band gap in insulating BLG}
The band gap of the finite temperature insulator as a function of temperature is calculated from the self consistent equation
\begin{equation}
	{\cal D}(T) = {\cal D}_{ext} + \delta{\cal D}_H(T,{\cal D}) + \delta{\cal D}_F(T,{\cal D}).
\end{equation}
 \begin{figure} [b!]
	\centering
	\includegraphics[width=0.84\linewidth]{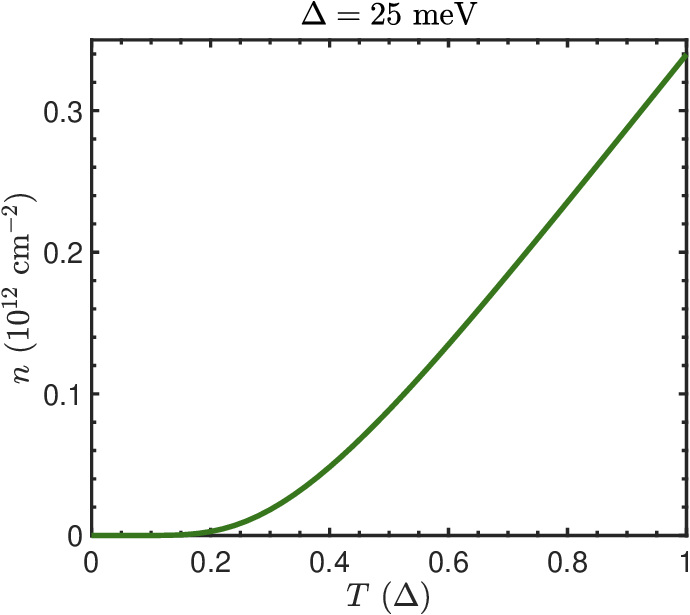}
	\caption{Density of thermally excited electrons versus temperature.
		In this plot the gap is kept constant at ${\cal D} = 2\Delta = 50$ meV} 
	\label{fig:n_T}
\end{figure}
Plots of ${\cal D}(T) $ vs temperature are shown in Fig.~\ref{fig:Delta_T}. Solid lines include both Hartree and Fock screening while dashed lines are for Hartree screening only. Red lines correspond to the external bias
${\cal D}_{ext}=50$ meV and blue lines correspond to ${\cal D}_{ext}=20$ meV. We show curves for suspended biased bilayer graphene ($\varepsilon=1$) and biased bilayer graphene encased in hBN ($\varepsilon=4$) with a metallic gate located at distance $d=50$ nm.

As stated in the previous section the Fock screening is quantitatively significant, changing the gap compared to its Hartree screened value by a factor of approximately $2$.
 Furthermore, at finite tempertature the Fock screening gives an interesting qualitative effect:
 a step-like reduction of the gap at $T\approx 0.2\Delta=0.1{\cal D}$.
 To explain this step-like reduction in Fig.~\ref{fig:n_T} we plot the density of thermally excited
 electrons versus temperature in Fig.~\ref{fig:n_T}. The density becomes sizable at $T>0.2\Delta$, below which it is essentially zero. 

 There is a slight change in the behaviour at $T\approx 0.2\Delta$ in the dashed Hartree lines as well,
 although this change in regime is much more subtle.

\section{Conclusions} \label{sec:Con}
We have developed the theory of Hartree-Fock screening of the band gap in biased bilayer graphene.
The screening comes from excess conduction electrons on the bilayer graphene. While Hartree screening has been studied previously, we consider the Fock screening and therefore the Hartree-Fock screening of the band gap in biased bilayer graphene for the first time.
We have shown that in the zero temperature insulating bilayer graphene account of the Fock screening
changes the gap by a factor of $\sim 2$ compared to the Hartree case.
Further, we have shown that in the zero tempeture metallic bilayer graphene the Fock contribution
changes the gap behaviour, resulting in a sharp reduction of the gap at very low density
of metallic charges.
Finally, we have considered the temperature dependence of the gap in insulating bilayer graphene.
Again, the Fock contribution changes the dependence qualitatively,  resulting in sharp step-like reduction
of the gap at low temperature.

\section{Acknowledgements} 
We acknowledge discussions with Dmitry Efimkin and Abhay Gupta. This work was supported by the Australian Research Council Centre of Excellence in Future Low-
Energy Electronics Technologies (CE170100039). 

\appendix
\section{$\Pi^{++}$ in the low momentum limit} \label{pi_approx}
In this appendix we present a derivation of the zero temperature component of polarization $\Pi^{++}$ in the limit $q\to0$. This is important to both validate numerical calculations of $\Pi^{++}$ and also to analytically investigate the behaviour of the Fock contribution to band gap renormalization in the small density limit. 

Starting from Eq.~\eqref{eq:pi}, we take the limit $q\to0$, where $\lim_{q\to 0} F^{++}_{p,p+q}=1$, $\lim_{q\to 0} (\epsilon_p-\epsilon_{p+q}) = \frac{\partial \epsilon}{\partial p}q \cos{\theta}$ and $\lim_{q\to 0}  n_p^+-n_{p+q}^+= \frac{\partial n}{\partial \epsilon}\frac{\partial \epsilon}{\partial p}q \cos{\theta}$.

\begin{equation}
	\lim_{q\to0}\Pi^{++} = \frac{1}{\pi^2}\int_0^{2\pi} \int_0^\infty p dp d\theta \frac{(\frac{\partial \epsilon}{\partial p})^2 q^2 \cos^2{\theta}\frac{\partial n}{\partial \epsilon}}{ \xi^2 + (\frac{\partial \epsilon}{\partial p})^2 q^2 \cos^2{\theta}}
\end{equation}

We make use of the relation $\frac{\partial n}{\partial \epsilon} = - \delta(\epsilon - \epsilon_F)$ and also make the substitution $\alpha = \frac{\xi}{\frac{\partial \epsilon}{\partial p}q}$.

\begin{equation}\label{eq:app1}
	\lim_{q\to0}\Pi^{++} = \frac{-1}{\pi^2}\int_0^{2\pi} \int_0^\infty \frac{d\theta \cos^2{\theta}}{\cos^2{\theta}+\alpha^2} p \delta(\epsilon - \epsilon_F) dp
\end{equation}

Let us consider the angular integral first. 

\begin{align}
	\int_0^{2\pi} \frac{d\theta \cos^2{\theta}}{\cos^2{\theta}+\alpha^2} &= \int_0^{2\pi} d\theta \left(1 - \frac{ \alpha^2}{\cos^2{\theta}+\alpha^2} \right) \nonumber\\ &= 2\pi - \alpha^2 I, \nonumber\\
	I &= \int_0^{2\pi} \frac{d\theta}{\cos^2{\theta}+\alpha^2}
\end{align}

We can shift the integral to the complex plane. Let $z = e^{2i\theta}$, $dz = 2izd\theta$. Note that we must integrate twice around the unit circle.

\begin{equation}
	I = \oint \frac{dz}{2iz(\frac{1}{4}(z+z^{-1} + 2)+\alpha^2} =  \frac{2}{i} \oint \frac{dz}{z^2+1+2z+4\alpha^2z}
\end{equation}

This integrand has simple poles $z = -(1+2\alpha^2) \pm 2\alpha\sqrt{\alpha^2+1}$. Only the pole $z = -(1+2\alpha^2) + 2\alpha\sqrt{\alpha^2+1}$ lies within the unit circle. Thus we use Cauchy's Residue Theorem and integrate twice around the unit circle to solve the integral.

\begin{equation}
	I = 2\pi\left(1-\frac{\alpha}{\sqrt{\alpha^2+1}}\right)
\end{equation}

Therefore with $dp= \frac{\partial p}{\partial \epsilon} d\epsilon=\frac{2m^2 \epsilon}{p^3}d\epsilon$ Eq.~\eqref{eq:app1} becomes

\begin{equation}
	\lim_{q\to0}\Pi^{++} = \frac{2}{\pi} \int \left(\frac{\xi}{\sqrt{\xi^2 + \frac{p^6}{4m^4 \epsilon^2} q^2}} -1\right) \frac{2m^2 \epsilon}{p^2}\delta(\epsilon - \epsilon_F) d\epsilon.
\end{equation}

Integrating over $\epsilon$ and noting $\epsilon_F\approx\Delta$ we find an expression for $\Pi{++}$ in the low momentum limit. 

\begin{equation}
	\lim_{q\to0}\Pi^{++} = -\frac{4 m^2\Delta }{\pi p_F^2} \left( 1- \frac{\xi }{\sqrt{\xi^2 + \frac{p_F^6 q^2}{4m^4\epsilon_F^2}}}\right)
\end{equation}

Now, it is also useful to consider the $\Pi^{++}$ in the limit $p_F\ll q\ll \sqrt{2m\Delta}$. For simplicity we consider the case $\xi=0$. In this regime $\epsilon_p-\epsilon_{p+q}\approx -q^4/8m^2\Delta$ and $n_{q+p}\approx0$, so Eq.~\eqref{eq:pi} gives

\begin{equation}
	\Pi^{++}(q,0) = 4\int_0^{p_F} \frac{pdp}{2\pi} \left(\frac{-8m^2\Delta}{q^4}\right), \; p_F\ll q \ll \sqrt{2m\Delta}.
\end{equation}

Straightforward integration yields 

\begin{equation} \label{pi_pp_app}
	\Pi^{++}(q,0) = \frac{-8m^2\Delta p_F^2}{\pi q^4}, \; p_F\ll q \ll \sqrt{2m\Delta}.
\end{equation}

\section{Low density Fock term}\label{low_density}
We start from Eq.~\eqref{eq:delta_m}. In the low density regime we may ignore the second term. Noting as $p_F \to 0$, $\Pi^{-+}\to 0$, we may rewrite the first term as
\begin{align}
	\delta {\cal D}_F = 4\int_0^\infty\int_0^\infty \frac{qdq}{2\pi}\frac{d\xi}{2\pi} \left( \frac{\Delta q^4/4m^2}{(\xi^2 + q^4/4m^2)^2 + 4\Delta^2\xi^2} \right)\nonumber\\ \times\frac{2\pi e^2}{\epsilon q - 2\pi e^2 \Pi^{++}(q,\xi)}.
\end{align}

Now, we are interested in the band gap renormalization due to the Fock term, which is calculated from the difference in the self energy between the doped and undoped case. We shall also perform the change of variable $\xi\to x$ defined by $\xi = xp^4/4m^2$.

\begin{align}
	\label{eq:app:b}
	&\delta {\cal D}_F(n)-\delta {\cal D}_F(n=0) \nonumber\\&= 4\int_0^\infty\int_0^\infty \frac{qdq}{2\pi}\frac{dx}{2\pi}\left( \frac{\Delta }{(1+x^2 q^4/4m^2)^2+4\Delta^2x^2} \right) \nonumber\\&\times\frac{2\pi e^2(2\pi e^2 \Pi^{++}(q,x))}{\epsilon q(\epsilon q - 2\pi e^2 \Pi^{++}(q,x))}
\end{align}

It is clear from analysis of the low density polarization $\Pi^{++}(q,x)$ presented in Appendix.\ref{pi_approx} that $\Pi^{++}(q,x)$ is very large in the regime of low $q$. Inspection of the integrand in Eq.~\eqref{eq:app:b} reveals that convergence of the integrand in $q$ space is determined by the second term, specifically

\begin{equation}
	\frac{2\pi e^2 \Pi^{++}(q,x))}{\epsilon q - 2\pi e^2 \Pi^{++}(q,x)}.
\end{equation}

From analysis of the first term in Eq.~\eqref{eq:app:b} we expect convergence in the $x$ domain around $x\sim 1/\Delta$, which remains small and does not greatly influence $\Pi^{++}(q,x)$. Therefore when performing $q$ space integration we may take polarization at $x=0$. We are interested in the small density limit, $p_F \ll q \ll \sqrt{2m\Delta}$, thus we take $\Pi^{++}(q,0)$ from Eq.~\eqref{pi_pp_app}.

\begin{align}
	&\delta {\cal D}_F(n)-\delta {\cal D}_F(n=0) \nonumber\\&= 4\int_0^\infty\int_0^\infty \frac{dq}{2\pi}\frac{dx}{2\pi}\left( \frac{\Delta }{(1+x^2 q^4/4m^2)^2+4\Delta^2x^2} \right) \nonumber\\&\times\frac{2\pi e^2(2\pi e^2 (-8m^2\Delta p_F^2/\pi q^4))}{\epsilon (\epsilon q + 2\pi e^2(8m^2\Delta p_F^2/\pi q^4))}
\end{align}

The integrand is convergent in q for $\epsilon q = 2\pi e^2(8m^2\Delta p_F^2/\pi q^4)$. Rearranging, and with $a_B = \epsilon/e^2m$ and $p_0=\sqrt{2m\Delta}$, the integrand is convergent at $q_{max} = \left[8p_0^2p_F^2/a_B\right]^{1/5}$. We may now evaluate the integral, noting $q_{max}\ll p_0$ so the $x^2q^4/4m^2$ term in the first term of the integral may be discarded.

\begin{align}
	&\delta {\cal D}_F(n)-\delta {\cal D}_F(n=0) \nonumber\\&\sim \frac{8\pi e^2}{\epsilon}\int_0^{q_{max}}\int_0^\infty \frac{dq}{2\pi}\frac{dx}{2\pi}\left( \frac{\Delta }{1+4\Delta^2x^2} \right)
	\nonumber\\
	&=\frac{e^2}{2\epsilon} \int_0^{q_{max}} dq
	\nonumber\\
	&=a_B Ry  \left(\frac{8p_0^2p_F^2}{a_B}\right)^{1/5}.
\end{align}

Here $Ry = me^4/2\epsilon^2$. With excess charge density given by $n=p_F^2/\pi$, it is clear that at low doping the Fock contribution to band gap renormalization scales as $n^{1/5}$.

\bibliography{BLG_BGR}

\begin{thebibliography}{60}%
\makeatletter
\providecommand \@ifxundefined [1]{%
 \@ifx{#1\undefined}
}%
\providecommand \@ifnum [1]{%
 \ifnum #1\expandafter \@firstoftwo
 \else \expandafter \@secondoftwo
 \fi
}%
\providecommand \@ifx [1]{%
 \ifx #1\expandafter \@firstoftwo
 \else \expandafter \@secondoftwo
 \fi
}%
\providecommand \natexlab [1]{#1}%
\providecommand \enquote  [1]{``#1''}%
\providecommand \bibnamefont  [1]{#1}%
\providecommand \bibfnamefont [1]{#1}%
\providecommand \citenamefont [1]{#1}%
\providecommand \href@noop [0]{\@secondoftwo}%
\providecommand \href [0]{\begingroup \@sanitize@url \@href}%
\providecommand \@href[1]{\@@startlink{#1}\@@href}%
\providecommand \@@href[1]{\endgroup#1\@@endlink}%
\providecommand \@sanitize@url [0]{\catcode `\\12\catcode `\$12\catcode
  `\&12\catcode `\#12\catcode `\^12\catcode `\_12\catcode `\%12\relax}%
\providecommand \@@startlink[1]{}%
\providecommand \@@endlink[0]{}%
\providecommand \url  [0]{\begingroup\@sanitize@url \@url }%
\providecommand \@url [1]{\endgroup\@href {#1}{\urlprefix }}%
\providecommand \urlprefix  [0]{URL }%
\providecommand \Eprint [0]{\href }%
\providecommand \doibase [0]{http://dx.doi.org/}%
\providecommand \selectlanguage [0]{\@gobble}%
\providecommand \bibinfo  [0]{\@secondoftwo}%
\providecommand \bibfield  [0]{\@secondoftwo}%
\providecommand \translation [1]{[#1]}%
\providecommand \BibitemOpen [0]{}%
\providecommand \bibitemStop [0]{}%
\providecommand \bibitemNoStop [0]{.\EOS\space}%
\providecommand \EOS [0]{\spacefactor3000\relax}%
\providecommand \BibitemShut  [1]{\csname bibitem#1\endcsname}%
\let\auto@bib@innerbib\@empty
\bibitem [{\citenamefont {Novoselov}\ \emph {et~al.}(2004)\citenamefont
  {Novoselov}, \citenamefont {Geim}, \citenamefont {Morozov}, \citenamefont
  {Jiang}, \citenamefont {Zhang}, \citenamefont {Dubonos}, \citenamefont
  {Grigorieva},\ and\ \citenamefont {Firsov}}]{novoselov_electric_2004}%
  \BibitemOpen
  \bibfield  {author} {\bibinfo {author} {\bibfnamefont {K.~S.}\ \bibnamefont
  {Novoselov}}, \bibinfo {author} {\bibfnamefont {A.~K.}\ \bibnamefont {Geim}},
  \bibinfo {author} {\bibfnamefont {S.~V.}\ \bibnamefont {Morozov}}, \bibinfo
  {author} {\bibfnamefont {D.}~\bibnamefont {Jiang}}, \bibinfo {author}
  {\bibfnamefont {Y.}~\bibnamefont {Zhang}}, \bibinfo {author} {\bibfnamefont
  {S.~V.}\ \bibnamefont {Dubonos}}, \bibinfo {author} {\bibfnamefont {I.~V.}\
  \bibnamefont {Grigorieva}}, \ and\ \bibinfo {author} {\bibfnamefont {A.~A.}\
  \bibnamefont {Firsov}},\ }\href {\doibase 10.1126/science.1102896} {\bibfield
   {journal} {\bibinfo  {journal} {Science}\ }\textbf {\bibinfo {volume}
  {306}},\ \bibinfo {pages} {666} (\bibinfo {year} {2004})}\BibitemShut
  {NoStop}%
\bibitem [{\citenamefont {McCann}\ and\ \citenamefont
  {Koshino}(2013)}]{mccann_electronic_2013}%
  \BibitemOpen
  \bibfield  {author} {\bibinfo {author} {\bibfnamefont {E.}~\bibnamefont
  {McCann}}\ and\ \bibinfo {author} {\bibfnamefont {M.}~\bibnamefont
  {Koshino}},\ }\href {\doibase 10.1088/0034-4885/76/5/056503} {\bibfield
  {journal} {\bibinfo  {journal} {Reports on Progress in Physics}\ }\textbf
  {\bibinfo {volume} {76}},\ \bibinfo {pages} {056503} (\bibinfo {year}
  {2013})}\BibitemShut {NoStop}%
\bibitem [{\citenamefont {Ohta}\ \emph {et~al.}(2006)\citenamefont {Ohta},
  \citenamefont {Bostwick}, \citenamefont {Seyller}, \citenamefont {Horn},\
  and\ \citenamefont {Rotenberg}}]{ohta_controlling_2006}%
  \BibitemOpen
  \bibfield  {author} {\bibinfo {author} {\bibfnamefont {T.}~\bibnamefont
  {Ohta}}, \bibinfo {author} {\bibfnamefont {A.}~\bibnamefont {Bostwick}},
  \bibinfo {author} {\bibfnamefont {T.}~\bibnamefont {Seyller}}, \bibinfo
  {author} {\bibfnamefont {K.}~\bibnamefont {Horn}}, \ and\ \bibinfo {author}
  {\bibfnamefont {E.}~\bibnamefont {Rotenberg}},\ }\href {\doibase
  10.1126/science.1130681} {\bibfield  {journal} {\bibinfo  {journal}
  {Science}\ }\textbf {\bibinfo {volume} {313}},\ \bibinfo {pages} {951}
  (\bibinfo {year} {2006})}\BibitemShut {NoStop}%
\bibitem [{\citenamefont {Castro}\ \emph {et~al.}(2007)\citenamefont {Castro},
  \citenamefont {Novoselov}, \citenamefont {Morozov}, \citenamefont {Peres},
  \citenamefont {Dos~Santos}, \citenamefont {Nilsson}, \citenamefont {Guinea},
  \citenamefont {Geim},\ and\ \citenamefont {Neto}}]{castro_biased_2007}%
  \BibitemOpen
  \bibfield  {author} {\bibinfo {author} {\bibfnamefont {E.~V.}\ \bibnamefont
  {Castro}}, \bibinfo {author} {\bibfnamefont {K.~S.}\ \bibnamefont
  {Novoselov}}, \bibinfo {author} {\bibfnamefont {S.~V.}\ \bibnamefont
  {Morozov}}, \bibinfo {author} {\bibfnamefont {N.~M.~R.}\ \bibnamefont
  {Peres}}, \bibinfo {author} {\bibfnamefont {J.~M. B.~L.}\ \bibnamefont
  {Dos~Santos}}, \bibinfo {author} {\bibfnamefont {J.}~\bibnamefont {Nilsson}},
  \bibinfo {author} {\bibfnamefont {F.}~\bibnamefont {Guinea}}, \bibinfo
  {author} {\bibfnamefont {A.~K.}\ \bibnamefont {Geim}}, \ and\ \bibinfo
  {author} {\bibfnamefont {A.~H.~C.}\ \bibnamefont {Neto}},\ }\href {\doibase
  10.1103/PhysRevLett.99.216802} {\bibfield  {journal} {\bibinfo  {journal}
  {Physical Review Letters}\ }\textbf {\bibinfo {volume} {99}},\ \bibinfo
  {pages} {216802} (\bibinfo {year} {2007})}\BibitemShut {NoStop}%
\bibitem [{\citenamefont {Zhou}\ \emph {et~al.}(2007)\citenamefont {Zhou},
  \citenamefont {Gweon}, \citenamefont {Fedorov}, \citenamefont {First},
  \citenamefont {De~Heer}, \citenamefont {Lee}, \citenamefont {Guinea},
  \citenamefont {Castro~Neto},\ and\ \citenamefont
  {Lanzara}}]{zhou_substrate-induced_2007}%
  \BibitemOpen
  \bibfield  {author} {\bibinfo {author} {\bibfnamefont {S.~Y.}\ \bibnamefont
  {Zhou}}, \bibinfo {author} {\bibfnamefont {G.-H.}\ \bibnamefont {Gweon}},
  \bibinfo {author} {\bibfnamefont {A.~V.}\ \bibnamefont {Fedorov}}, \bibinfo
  {author} {\bibfnamefont {P.~N.}\ \bibnamefont {First}}, \bibinfo {author}
  {\bibfnamefont {W.~A.}\ \bibnamefont {De~Heer}}, \bibinfo {author}
  {\bibfnamefont {D.-H.}\ \bibnamefont {Lee}}, \bibinfo {author} {\bibfnamefont
  {F.}~\bibnamefont {Guinea}}, \bibinfo {author} {\bibfnamefont {A.~H.}\
  \bibnamefont {Castro~Neto}}, \ and\ \bibinfo {author} {\bibfnamefont
  {A.}~\bibnamefont {Lanzara}},\ }\href {\doibase 10.1038/nmat2003} {\bibfield
  {journal} {\bibinfo  {journal} {Nature Materials}\ }\textbf {\bibinfo
  {volume} {6}},\ \bibinfo {pages} {770} (\bibinfo {year} {2007})}\BibitemShut
  {NoStop}%
\bibitem [{\citenamefont {Alattas}\ and\ \citenamefont
  {Schwingenschlögl}(2018)}]{alattas_band_2018}%
  \BibitemOpen
  \bibfield  {author} {\bibinfo {author} {\bibfnamefont {M.}~\bibnamefont
  {Alattas}}\ and\ \bibinfo {author} {\bibfnamefont {U.}~\bibnamefont
  {Schwingenschlögl}},\ }\href {\doibase 10.1038/s41598-018-35671-2}
  {\bibfield  {journal} {\bibinfo  {journal} {Scientific Reports}\ }\textbf
  {\bibinfo {volume} {8}},\ \bibinfo {pages} {17689} (\bibinfo {year}
  {2018})}\BibitemShut {NoStop}%
\bibitem [{\citenamefont {Zhang}\ \emph {et~al.}(2009)\citenamefont {Zhang},
  \citenamefont {Tang}, \citenamefont {Girit}, \citenamefont {Hao},
  \citenamefont {Martin}, \citenamefont {Zettl}, \citenamefont {Crommie},
  \citenamefont {Shen},\ and\ \citenamefont {Wang}}]{zhang_direct_2009}%
  \BibitemOpen
  \bibfield  {author} {\bibinfo {author} {\bibfnamefont {Y.}~\bibnamefont
  {Zhang}}, \bibinfo {author} {\bibfnamefont {T.-T.}\ \bibnamefont {Tang}},
  \bibinfo {author} {\bibfnamefont {C.}~\bibnamefont {Girit}}, \bibinfo
  {author} {\bibfnamefont {Z.}~\bibnamefont {Hao}}, \bibinfo {author}
  {\bibfnamefont {M.~C.}\ \bibnamefont {Martin}}, \bibinfo {author}
  {\bibfnamefont {A.}~\bibnamefont {Zettl}}, \bibinfo {author} {\bibfnamefont
  {M.~F.}\ \bibnamefont {Crommie}}, \bibinfo {author} {\bibfnamefont {Y.~R.}\
  \bibnamefont {Shen}}, \ and\ \bibinfo {author} {\bibfnamefont
  {F.}~\bibnamefont {Wang}},\ }\href {\doibase 10.1038/nature08105} {\bibfield
  {journal} {\bibinfo  {journal} {Nature}\ }\textbf {\bibinfo {volume} {459}},\
  \bibinfo {pages} {820} (\bibinfo {year} {2009})}\BibitemShut {NoStop}%
\bibitem [{\citenamefont {Guinea}\ \emph {et~al.}(2006)\citenamefont {Guinea},
  \citenamefont {Castro~Neto},\ and\ \citenamefont
  {Peres}}]{guinea_electronic_2006}%
  \BibitemOpen
  \bibfield  {author} {\bibinfo {author} {\bibfnamefont {F.}~\bibnamefont
  {Guinea}}, \bibinfo {author} {\bibfnamefont {A.~H.}\ \bibnamefont
  {Castro~Neto}}, \ and\ \bibinfo {author} {\bibfnamefont {N.~M.~R.}\
  \bibnamefont {Peres}},\ }\href {\doibase 10.1103/PhysRevB.73.245426}
  {\bibfield  {journal} {\bibinfo  {journal} {Physical Review B}\ }\textbf
  {\bibinfo {volume} {73}},\ \bibinfo {pages} {245426} (\bibinfo {year}
  {2006})}\BibitemShut {NoStop}%
\bibitem [{\citenamefont {Oostinga}\ \emph {et~al.}(2008)\citenamefont
  {Oostinga}, \citenamefont {Heersche}, \citenamefont {Liu}, \citenamefont
  {Morpurgo},\ and\ \citenamefont {Vandersypen}}]{oostinga_gate-induced_2008}%
  \BibitemOpen
  \bibfield  {author} {\bibinfo {author} {\bibfnamefont {J.~B.}\ \bibnamefont
  {Oostinga}}, \bibinfo {author} {\bibfnamefont {H.~B.}\ \bibnamefont
  {Heersche}}, \bibinfo {author} {\bibfnamefont {X.}~\bibnamefont {Liu}},
  \bibinfo {author} {\bibfnamefont {A.~F.}\ \bibnamefont {Morpurgo}}, \ and\
  \bibinfo {author} {\bibfnamefont {L.~M.~K.}\ \bibnamefont {Vandersypen}},\
  }\href {\doibase 10.1038/nmat2082} {\bibfield  {journal} {\bibinfo  {journal}
  {Nature Materials}\ }\textbf {\bibinfo {volume} {7}},\ \bibinfo {pages} {151}
  (\bibinfo {year} {2008})}\BibitemShut {NoStop}%
\bibitem [{\citenamefont {Anderson}\ \emph {et~al.}(2023)\citenamefont
  {Anderson}, \citenamefont {Natera-Cordero}, \citenamefont
  {Guarochico-Moreira}, \citenamefont {Grigorieva},\ and\ \citenamefont
  {Vera-Marun}}]{anderson_exploring_2023}%
  \BibitemOpen
  \bibfield  {author} {\bibinfo {author} {\bibfnamefont {C.~R.}\ \bibnamefont
  {Anderson}}, \bibinfo {author} {\bibfnamefont {N.}~\bibnamefont
  {Natera-Cordero}}, \bibinfo {author} {\bibfnamefont {V.~H.}\ \bibnamefont
  {Guarochico-Moreira}}, \bibinfo {author} {\bibfnamefont {I.~V.}\ \bibnamefont
  {Grigorieva}}, \ and\ \bibinfo {author} {\bibfnamefont {I.~J.}\ \bibnamefont
  {Vera-Marun}},\ }\href {\doibase 10.1038/s41598-023-36800-2} {\bibfield
  {journal} {\bibinfo  {journal} {Scientific Reports}\ }\textbf {\bibinfo
  {volume} {13}},\ \bibinfo {pages} {10343} (\bibinfo {year}
  {2023})}\BibitemShut {NoStop}%
\bibitem [{\citenamefont {Yang}(2011)}]{yang_excitons_2011}%
  \BibitemOpen
  \bibfield  {author} {\bibinfo {author} {\bibfnamefont {L.}~\bibnamefont
  {Yang}},\ }\href {\doibase 10.1103/PhysRevB.83.085405} {\bibfield  {journal}
  {\bibinfo  {journal} {Physical Review B}\ }\textbf {\bibinfo {volume} {83}},\
  \bibinfo {pages} {085405} (\bibinfo {year} {2011})}\BibitemShut {NoStop}%
\bibitem [{\citenamefont {Ju}\ \emph {et~al.}(2017)\citenamefont {Ju},
  \citenamefont {Wang}, \citenamefont {Cao}, \citenamefont {Taniguchi},
  \citenamefont {Watanabe}, \citenamefont {Louie}, \citenamefont {Rana},
  \citenamefont {Park}, \citenamefont {Hone}, \citenamefont {Wang},\ and\
  \citenamefont {McEuen}}]{ju_tunable_2017}%
  \BibitemOpen
  \bibfield  {author} {\bibinfo {author} {\bibfnamefont {L.}~\bibnamefont
  {Ju}}, \bibinfo {author} {\bibfnamefont {L.}~\bibnamefont {Wang}}, \bibinfo
  {author} {\bibfnamefont {T.}~\bibnamefont {Cao}}, \bibinfo {author}
  {\bibfnamefont {T.}~\bibnamefont {Taniguchi}}, \bibinfo {author}
  {\bibfnamefont {K.}~\bibnamefont {Watanabe}}, \bibinfo {author}
  {\bibfnamefont {S.~G.}\ \bibnamefont {Louie}}, \bibinfo {author}
  {\bibfnamefont {F.}~\bibnamefont {Rana}}, \bibinfo {author} {\bibfnamefont
  {J.}~\bibnamefont {Park}}, \bibinfo {author} {\bibfnamefont {J.}~\bibnamefont
  {Hone}}, \bibinfo {author} {\bibfnamefont {F.}~\bibnamefont {Wang}}, \ and\
  \bibinfo {author} {\bibfnamefont {P.~L.}\ \bibnamefont {McEuen}},\ }\href
  {\doibase 10.1126/science.aam9175} {\bibfield  {journal} {\bibinfo  {journal}
  {Science}\ }\textbf {\bibinfo {volume} {358}},\ \bibinfo {pages} {907}
  (\bibinfo {year} {2017})}\BibitemShut {NoStop}%
\bibitem [{\citenamefont {Li}\ \emph {et~al.}(2017)\citenamefont {Li},
  \citenamefont {Taniguchi}, \citenamefont {Watanabe}, \citenamefont {Hone},\
  and\ \citenamefont {Dean}}]{li_excitonic_2017}%
  \BibitemOpen
  \bibfield  {author} {\bibinfo {author} {\bibfnamefont {J.~I.~A.}\
  \bibnamefont {Li}}, \bibinfo {author} {\bibfnamefont {T.}~\bibnamefont
  {Taniguchi}}, \bibinfo {author} {\bibfnamefont {K.}~\bibnamefont {Watanabe}},
  \bibinfo {author} {\bibfnamefont {J.}~\bibnamefont {Hone}}, \ and\ \bibinfo
  {author} {\bibfnamefont {C.~R.}\ \bibnamefont {Dean}},\ }\href {\doibase
  10.1038/nphys4140} {\bibfield  {journal} {\bibinfo  {journal} {Nature
  Physics}\ }\textbf {\bibinfo {volume} {13}},\ \bibinfo {pages} {751}
  (\bibinfo {year} {2017})}\BibitemShut {NoStop}%
\bibitem [{\citenamefont {Henriques}\ \emph {et~al.}(2022)\citenamefont
  {Henriques}, \citenamefont {Epstein},\ and\ \citenamefont
  {Peres}}]{henriques_absorption_2022}%
  \BibitemOpen
  \bibfield  {author} {\bibinfo {author} {\bibfnamefont {J.~C.~G.}\
  \bibnamefont {Henriques}}, \bibinfo {author} {\bibfnamefont {I.}~\bibnamefont
  {Epstein}}, \ and\ \bibinfo {author} {\bibfnamefont {N.~M.~R.}\ \bibnamefont
  {Peres}},\ }\href {\doibase 10.1103/PhysRevB.105.045411} {\bibfield
  {journal} {\bibinfo  {journal} {Physical Review B}\ }\textbf {\bibinfo
  {volume} {105}},\ \bibinfo {pages} {045411} (\bibinfo {year}
  {2022})}\BibitemShut {NoStop}%
\bibitem [{\citenamefont {Sauer}\ and\ \citenamefont
  {Pedersen}(2022)}]{sauer_exciton_2022}%
  \BibitemOpen
  \bibfield  {author} {\bibinfo {author} {\bibfnamefont {M.~O.}\ \bibnamefont
  {Sauer}}\ and\ \bibinfo {author} {\bibfnamefont {T.~G.}\ \bibnamefont
  {Pedersen}},\ }\href {\doibase 10.1103/PhysRevB.105.115416} {\bibfield
  {journal} {\bibinfo  {journal} {Physical Review B}\ }\textbf {\bibinfo
  {volume} {105}},\ \bibinfo {pages} {115416} (\bibinfo {year}
  {2022})}\BibitemShut {NoStop}%
\bibitem [{\citenamefont {Liu}\ \emph {et~al.}(2023)\citenamefont {Liu},
  \citenamefont {Annawati}, \citenamefont {Hung}, \citenamefont {Gulo},
  \citenamefont {Solís-Fernández}, \citenamefont {Kawahara}, \citenamefont
  {Ago},\ and\ \citenamefont {Saito}}]{liu_interference_2023}%
  \BibitemOpen
  \bibfield  {author} {\bibinfo {author} {\bibfnamefont {H.-L.}\ \bibnamefont
  {Liu}}, \bibinfo {author} {\bibfnamefont {B.~D.}\ \bibnamefont {Annawati}},
  \bibinfo {author} {\bibfnamefont {N.~T.}\ \bibnamefont {Hung}}, \bibinfo
  {author} {\bibfnamefont {D.~P.}\ \bibnamefont {Gulo}}, \bibinfo {author}
  {\bibfnamefont {P.}~\bibnamefont {Solís-Fernández}}, \bibinfo {author}
  {\bibfnamefont {K.}~\bibnamefont {Kawahara}}, \bibinfo {author}
  {\bibfnamefont {H.}~\bibnamefont {Ago}}, \ and\ \bibinfo {author}
  {\bibfnamefont {R.}~\bibnamefont {Saito}},\ }\href {\doibase
  10.1103/PhysRevB.107.165421} {\bibfield  {journal} {\bibinfo  {journal}
  {Physical Review B}\ }\textbf {\bibinfo {volume} {107}},\ \bibinfo {pages}
  {165421} (\bibinfo {year} {2023})}\BibitemShut {NoStop}%
\bibitem [{\citenamefont {Saleem}\ \emph {et~al.}(2023)\citenamefont {Saleem},
  \citenamefont {Sadecka}, \citenamefont {Korkusinski}, \citenamefont
  {Miravet}, \citenamefont {Dusko},\ and\ \citenamefont
  {Hawrylak}}]{saleem_theory_2023}%
  \BibitemOpen
  \bibfield  {author} {\bibinfo {author} {\bibfnamefont {Y.}~\bibnamefont
  {Saleem}}, \bibinfo {author} {\bibfnamefont {K.}~\bibnamefont {Sadecka}},
  \bibinfo {author} {\bibfnamefont {M.}~\bibnamefont {Korkusinski}}, \bibinfo
  {author} {\bibfnamefont {D.}~\bibnamefont {Miravet}}, \bibinfo {author}
  {\bibfnamefont {A.}~\bibnamefont {Dusko}}, \ and\ \bibinfo {author}
  {\bibfnamefont {P.}~\bibnamefont {Hawrylak}},\ }\href {\doibase
  10.1021/acs.nanolett.3c00406} {\bibfield  {journal} {\bibinfo  {journal}
  {Nano Letters}\ }\textbf {\bibinfo {volume} {23}},\ \bibinfo {pages} {2998}
  (\bibinfo {year} {2023})}\BibitemShut {NoStop}%
\bibitem [{\citenamefont {Scammell}\ and\ \citenamefont
  {Sushkov}(2023{\natexlab{a}})}]{scammell_dynamical_2023}%
  \BibitemOpen
  \bibfield  {author} {\bibinfo {author} {\bibfnamefont {H.~D.}\ \bibnamefont
  {Scammell}}\ and\ \bibinfo {author} {\bibfnamefont {O.~P.}\ \bibnamefont
  {Sushkov}},\ }\href {\doibase 10.1103/PhysRevB.107.085104} {\bibfield
  {journal} {\bibinfo  {journal} {Physical Review B}\ }\textbf {\bibinfo
  {volume} {107}},\ \bibinfo {pages} {085104} (\bibinfo {year}
  {2023}{\natexlab{a}})}\BibitemShut {NoStop}%
\bibitem [{\citenamefont {Scammell}\ and\ \citenamefont
  {Sushkov}(2023{\natexlab{b}})}]{scammell_exciton_2023}%
  \BibitemOpen
  \bibfield  {author} {\bibinfo {author} {\bibfnamefont {H.~D.}\ \bibnamefont
  {Scammell}}\ and\ \bibinfo {author} {\bibfnamefont {O.~P.}\ \bibnamefont
  {Sushkov}},\ }\href {\doibase 10.1103/PhysRevResearch.5.043176} {\bibfield
  {journal} {\bibinfo  {journal} {Physical Review Research}\ }\textbf {\bibinfo
  {volume} {5}},\ \bibinfo {pages} {043176} (\bibinfo {year}
  {2023}{\natexlab{b}})}\BibitemShut {NoStop}%
\bibitem [{\citenamefont {Gusev}\ \emph {et~al.}(2021)\citenamefont {Gusev},
  \citenamefont {Jaroshevich}, \citenamefont {Levin}, \citenamefont {Kvon},\
  and\ \citenamefont {Bakarov}}]{gusev_viscous_2021}%
  \BibitemOpen
  \bibfield  {author} {\bibinfo {author} {\bibfnamefont {G.~M.}\ \bibnamefont
  {Gusev}}, \bibinfo {author} {\bibfnamefont {A.~S.}\ \bibnamefont
  {Jaroshevich}}, \bibinfo {author} {\bibfnamefont {A.~D.}\ \bibnamefont
  {Levin}}, \bibinfo {author} {\bibfnamefont {Z.~D.}\ \bibnamefont {Kvon}}, \
  and\ \bibinfo {author} {\bibfnamefont {A.~K.}\ \bibnamefont {Bakarov}},\
  }\href {\doibase 10.1103/PhysRevB.103.075303} {\bibfield  {journal} {\bibinfo
   {journal} {Physical Review B}\ }\textbf {\bibinfo {volume} {103}},\ \bibinfo
  {pages} {075303} (\bibinfo {year} {2021})}\BibitemShut {NoStop}%
\bibitem [{\citenamefont {Mönch}\ \emph {et~al.}(2022)\citenamefont {Mönch},
  \citenamefont {Potashin}, \citenamefont {Lindner}, \citenamefont {Yahniuk},
  \citenamefont {Golub}, \citenamefont {Kachorovskii}, \citenamefont {Bel'kov},
  \citenamefont {Huber}, \citenamefont {Watanabe}, \citenamefont {Taniguchi},
  \citenamefont {Eroms}, \citenamefont {Weiss},\ and\ \citenamefont
  {Ganichev}}]{monch_ratchet_2022}%
  \BibitemOpen
  \bibfield  {author} {\bibinfo {author} {\bibfnamefont {E.}~\bibnamefont
  {Mönch}}, \bibinfo {author} {\bibfnamefont {S.~O.}\ \bibnamefont
  {Potashin}}, \bibinfo {author} {\bibfnamefont {K.}~\bibnamefont {Lindner}},
  \bibinfo {author} {\bibfnamefont {I.}~\bibnamefont {Yahniuk}}, \bibinfo
  {author} {\bibfnamefont {L.~E.}\ \bibnamefont {Golub}}, \bibinfo {author}
  {\bibfnamefont {V.~Y.}\ \bibnamefont {Kachorovskii}}, \bibinfo {author}
  {\bibfnamefont {V.~V.}\ \bibnamefont {Bel'kov}}, \bibinfo {author}
  {\bibfnamefont {R.}~\bibnamefont {Huber}}, \bibinfo {author} {\bibfnamefont
  {K.}~\bibnamefont {Watanabe}}, \bibinfo {author} {\bibfnamefont
  {T.}~\bibnamefont {Taniguchi}}, \bibinfo {author} {\bibfnamefont
  {J.}~\bibnamefont {Eroms}}, \bibinfo {author} {\bibfnamefont
  {D.}~\bibnamefont {Weiss}}, \ and\ \bibinfo {author} {\bibfnamefont {S.~D.}\
  \bibnamefont {Ganichev}},\ }\href {\doibase 10.1103/PhysRevB.105.045404}
  {\bibfield  {journal} {\bibinfo  {journal} {Physical Review B}\ }\textbf
  {\bibinfo {volume} {105}},\ \bibinfo {pages} {045404} (\bibinfo {year}
  {2022})}\BibitemShut {NoStop}%
\bibitem [{\citenamefont {Cruise}\ \emph {et~al.}(2024)\citenamefont {Cruise},
  \citenamefont {Seidel}, \citenamefont {Henriksen},\ and\ \citenamefont
  {Vignale}}]{cruise_observability_2024}%
  \BibitemOpen
  \bibfield  {author} {\bibinfo {author} {\bibfnamefont {J.~R.}\ \bibnamefont
  {Cruise}}, \bibinfo {author} {\bibfnamefont {A.}~\bibnamefont {Seidel}},
  \bibinfo {author} {\bibfnamefont {E.~A.}\ \bibnamefont {Henriksen}}, \ and\
  \bibinfo {author} {\bibfnamefont {G.}~\bibnamefont {Vignale}},\ }\href
  {\doibase 10.1103/PhysRevB.110.115416} {\bibfield  {journal} {\bibinfo
  {journal} {Physical Review B}\ }\textbf {\bibinfo {volume} {110}},\ \bibinfo
  {pages} {115416} (\bibinfo {year} {2024})}\BibitemShut {NoStop}%
\bibitem [{\citenamefont {Grigorenko}\ \emph {et~al.}(2012)\citenamefont
  {Grigorenko}, \citenamefont {Polini},\ and\ \citenamefont
  {Novoselov}}]{grigorenko_graphene_2012}%
  \BibitemOpen
  \bibfield  {author} {\bibinfo {author} {\bibfnamefont {A.~N.}\ \bibnamefont
  {Grigorenko}}, \bibinfo {author} {\bibfnamefont {M.}~\bibnamefont {Polini}},
  \ and\ \bibinfo {author} {\bibfnamefont {K.~S.}\ \bibnamefont {Novoselov}},\
  }\href {\doibase 10.1038/nphoton.2012.262} {\bibfield  {journal} {\bibinfo
  {journal} {Nature Photonics}\ }\textbf {\bibinfo {volume} {6}},\ \bibinfo
  {pages} {749} (\bibinfo {year} {2012})}\BibitemShut {NoStop}%
\bibitem [{\citenamefont {Fei}\ \emph {et~al.}(2015)\citenamefont {Fei},
  \citenamefont {Iwinski}, \citenamefont {Ni}, \citenamefont {Zhang},
  \citenamefont {Bao}, \citenamefont {Rodin}, \citenamefont {Lee},
  \citenamefont {Wagner}, \citenamefont {Liu}, \citenamefont {Dai},
  \citenamefont {Goldflam}, \citenamefont {Thiemens}, \citenamefont {Keilmann},
  \citenamefont {Lau}, \citenamefont {Castro-Neto}, \citenamefont {Fogler},\
  and\ \citenamefont {Basov}}]{fei_tunneling_2015}%
  \BibitemOpen
  \bibfield  {author} {\bibinfo {author} {\bibfnamefont {Z.}~\bibnamefont
  {Fei}}, \bibinfo {author} {\bibfnamefont {E.~G.}\ \bibnamefont {Iwinski}},
  \bibinfo {author} {\bibfnamefont {G.~X.}\ \bibnamefont {Ni}}, \bibinfo
  {author} {\bibfnamefont {L.~M.}\ \bibnamefont {Zhang}}, \bibinfo {author}
  {\bibfnamefont {W.}~\bibnamefont {Bao}}, \bibinfo {author} {\bibfnamefont
  {A.~S.}\ \bibnamefont {Rodin}}, \bibinfo {author} {\bibfnamefont
  {Y.}~\bibnamefont {Lee}}, \bibinfo {author} {\bibfnamefont {M.}~\bibnamefont
  {Wagner}}, \bibinfo {author} {\bibfnamefont {M.~K.}\ \bibnamefont {Liu}},
  \bibinfo {author} {\bibfnamefont {S.}~\bibnamefont {Dai}}, \bibinfo {author}
  {\bibfnamefont {M.~D.}\ \bibnamefont {Goldflam}}, \bibinfo {author}
  {\bibfnamefont {M.}~\bibnamefont {Thiemens}}, \bibinfo {author}
  {\bibfnamefont {F.}~\bibnamefont {Keilmann}}, \bibinfo {author}
  {\bibfnamefont {C.~N.}\ \bibnamefont {Lau}}, \bibinfo {author} {\bibfnamefont
  {A.~H.}\ \bibnamefont {Castro-Neto}}, \bibinfo {author} {\bibfnamefont
  {M.~M.}\ \bibnamefont {Fogler}}, \ and\ \bibinfo {author} {\bibfnamefont
  {D.~N.}\ \bibnamefont {Basov}},\ }\href {\doibase
  10.1021/acs.nanolett.5b00912} {\bibfield  {journal} {\bibinfo  {journal}
  {Nano Letters}\ }\textbf {\bibinfo {volume} {15}},\ \bibinfo {pages} {4973}
  (\bibinfo {year} {2015})}\BibitemShut {NoStop}%
\bibitem [{\citenamefont {Anirban}(2021)}]{anirban_superconductivity_2021}%
  \BibitemOpen
  \bibfield  {author} {\bibinfo {author} {\bibfnamefont {A.}~\bibnamefont
  {Anirban}},\ }\href {\doibase 10.1038/s42254-021-00413-3} {\bibfield
  {journal} {\bibinfo  {journal} {Nature Reviews Physics}\ }\textbf {\bibinfo
  {volume} {4}},\ \bibinfo {pages} {8} (\bibinfo {year} {2021})}\BibitemShut
  {NoStop}%
\bibitem [{\citenamefont {Li}\ \emph {et~al.}(2024)\citenamefont {Li},
  \citenamefont {Xu}, \citenamefont {Li}, \citenamefont {Li}, \citenamefont
  {Li}, \citenamefont {Watanabe}, \citenamefont {Taniguchi}, \citenamefont
  {Tong}, \citenamefont {Shen}, \citenamefont {Lu}, \citenamefont {Jia},
  \citenamefont {Wu}, \citenamefont {Liu},\ and\ \citenamefont
  {Li}}]{li_tunable_2024}%
  \BibitemOpen
  \bibfield  {author} {\bibinfo {author} {\bibfnamefont {C.}~\bibnamefont
  {Li}}, \bibinfo {author} {\bibfnamefont {F.}~\bibnamefont {Xu}}, \bibinfo
  {author} {\bibfnamefont {B.}~\bibnamefont {Li}}, \bibinfo {author}
  {\bibfnamefont {J.}~\bibnamefont {Li}}, \bibinfo {author} {\bibfnamefont
  {G.}~\bibnamefont {Li}}, \bibinfo {author} {\bibfnamefont {K.}~\bibnamefont
  {Watanabe}}, \bibinfo {author} {\bibfnamefont {T.}~\bibnamefont {Taniguchi}},
  \bibinfo {author} {\bibfnamefont {B.}~\bibnamefont {Tong}}, \bibinfo {author}
  {\bibfnamefont {J.}~\bibnamefont {Shen}}, \bibinfo {author} {\bibfnamefont
  {L.}~\bibnamefont {Lu}}, \bibinfo {author} {\bibfnamefont {J.}~\bibnamefont
  {Jia}}, \bibinfo {author} {\bibfnamefont {F.}~\bibnamefont {Wu}}, \bibinfo
  {author} {\bibfnamefont {X.}~\bibnamefont {Liu}}, \ and\ \bibinfo {author}
  {\bibfnamefont {T.}~\bibnamefont {Li}},\ }\href {\doibase
  10.1038/s41586-024-07584-w} {\bibfield  {journal} {\bibinfo  {journal}
  {Nature}\ }\textbf {\bibinfo {volume} {631}},\ \bibinfo {pages} {300}
  (\bibinfo {year} {2024})}\BibitemShut {NoStop}%
\bibitem [{\citenamefont {Wagner}\ \emph {et~al.}(2024)\citenamefont {Wagner},
  \citenamefont {Kwan}, \citenamefont {Bultinck}, \citenamefont {Simon},\ and\
  \citenamefont {Parameswaran}}]{wagner_superconductivity_2024}%
  \BibitemOpen
  \bibfield  {author} {\bibinfo {author} {\bibfnamefont {G.}~\bibnamefont
  {Wagner}}, \bibinfo {author} {\bibfnamefont {Y.~H.}\ \bibnamefont {Kwan}},
  \bibinfo {author} {\bibfnamefont {N.}~\bibnamefont {Bultinck}}, \bibinfo
  {author} {\bibfnamefont {S.~H.}\ \bibnamefont {Simon}}, \ and\ \bibinfo
  {author} {\bibfnamefont {S.~A.}\ \bibnamefont {Parameswaran}},\ }\href
  {\doibase 10.1103/PhysRevB.110.214517} {\bibfield  {journal} {\bibinfo
  {journal} {Physical Review B}\ }\textbf {\bibinfo {volume} {110}},\ \bibinfo
  {pages} {214517} (\bibinfo {year} {2024})}\BibitemShut {NoStop}%
\bibitem [{\citenamefont {Sboychakov}\ \emph {et~al.}(2023)\citenamefont
  {Sboychakov}, \citenamefont {Rozhkov},\ and\ \citenamefont
  {Rakhmanov}}]{sboychakov_triplet_2023}%
  \BibitemOpen
  \bibfield  {author} {\bibinfo {author} {\bibfnamefont {A.~O.}\ \bibnamefont
  {Sboychakov}}, \bibinfo {author} {\bibfnamefont {A.~V.}\ \bibnamefont
  {Rozhkov}}, \ and\ \bibinfo {author} {\bibfnamefont {A.~L.}\ \bibnamefont
  {Rakhmanov}},\ }\href {\doibase 10.1103/PhysRevB.108.184503} {\bibfield
  {journal} {\bibinfo  {journal} {Physical Review B}\ }\textbf {\bibinfo
  {volume} {108}},\ \bibinfo {pages} {184503} (\bibinfo {year}
  {2023})}\BibitemShut {NoStop}%
\bibitem [{\citenamefont {Zhou}\ \emph {et~al.}(2022)\citenamefont {Zhou},
  \citenamefont {Holleis}, \citenamefont {Saito}, \citenamefont {Cohen},
  \citenamefont {Huynh}, \citenamefont {Patterson}, \citenamefont {Yang},
  \citenamefont {Taniguchi}, \citenamefont {Watanabe},\ and\ \citenamefont
  {Young}}]{zhou_isospin_2022}%
  \BibitemOpen
  \bibfield  {author} {\bibinfo {author} {\bibfnamefont {H.}~\bibnamefont
  {Zhou}}, \bibinfo {author} {\bibfnamefont {L.}~\bibnamefont {Holleis}},
  \bibinfo {author} {\bibfnamefont {Y.}~\bibnamefont {Saito}}, \bibinfo
  {author} {\bibfnamefont {L.}~\bibnamefont {Cohen}}, \bibinfo {author}
  {\bibfnamefont {W.}~\bibnamefont {Huynh}}, \bibinfo {author} {\bibfnamefont
  {C.~L.}\ \bibnamefont {Patterson}}, \bibinfo {author} {\bibfnamefont
  {F.}~\bibnamefont {Yang}}, \bibinfo {author} {\bibfnamefont {T.}~\bibnamefont
  {Taniguchi}}, \bibinfo {author} {\bibfnamefont {K.}~\bibnamefont {Watanabe}},
  \ and\ \bibinfo {author} {\bibfnamefont {A.~F.}\ \bibnamefont {Young}},\
  }\href {\doibase 10.1126/science.abm8386} {\bibfield  {journal} {\bibinfo
  {journal} {Science}\ }\textbf {\bibinfo {volume} {375}},\ \bibinfo {pages}
  {774} (\bibinfo {year} {2022})}\BibitemShut {NoStop}%
\bibitem [{\citenamefont {Cao}\ \emph {et~al.}(2018)\citenamefont {Cao},
  \citenamefont {Fatemi}, \citenamefont {Fang}, \citenamefont {Watanabe},
  \citenamefont {Taniguchi}, \citenamefont {Kaxiras},\ and\ \citenamefont
  {Jarillo-Herrero}}]{cao_unconventional_2018}%
  \BibitemOpen
  \bibfield  {author} {\bibinfo {author} {\bibfnamefont {Y.}~\bibnamefont
  {Cao}}, \bibinfo {author} {\bibfnamefont {V.}~\bibnamefont {Fatemi}},
  \bibinfo {author} {\bibfnamefont {S.}~\bibnamefont {Fang}}, \bibinfo {author}
  {\bibfnamefont {K.}~\bibnamefont {Watanabe}}, \bibinfo {author}
  {\bibfnamefont {T.}~\bibnamefont {Taniguchi}}, \bibinfo {author}
  {\bibfnamefont {E.}~\bibnamefont {Kaxiras}}, \ and\ \bibinfo {author}
  {\bibfnamefont {P.}~\bibnamefont {Jarillo-Herrero}},\ }\href {\doibase
  10.1038/nature26160} {\bibfield  {journal} {\bibinfo  {journal} {Nature}\
  }\textbf {\bibinfo {volume} {556}},\ \bibinfo {pages} {43} (\bibinfo {year}
  {2018})}\BibitemShut {NoStop}%
\bibitem [{\citenamefont {Sharma}\ \emph {et~al.}(2020)\citenamefont {Sharma},
  \citenamefont {Trushin}, \citenamefont {Sushkov}, \citenamefont {Vignale},\
  and\ \citenamefont {Adam}}]{sharma_superconductivity_2020}%
  \BibitemOpen
  \bibfield  {author} {\bibinfo {author} {\bibfnamefont {G.}~\bibnamefont
  {Sharma}}, \bibinfo {author} {\bibfnamefont {M.}~\bibnamefont {Trushin}},
  \bibinfo {author} {\bibfnamefont {O.~P.}\ \bibnamefont {Sushkov}}, \bibinfo
  {author} {\bibfnamefont {G.}~\bibnamefont {Vignale}}, \ and\ \bibinfo
  {author} {\bibfnamefont {S.}~\bibnamefont {Adam}},\ }\href {\doibase
  10.1103/PhysRevResearch.2.022040} {\bibfield  {journal} {\bibinfo  {journal}
  {Physical Review Research}\ }\textbf {\bibinfo {volume} {2}},\ \bibinfo
  {pages} {022040} (\bibinfo {year} {2020})}\BibitemShut {NoStop}%
\bibitem [{\citenamefont {Cao}\ \emph {et~al.}(2021)\citenamefont {Cao},
  \citenamefont {Rodan-Legrain}, \citenamefont {Park}, \citenamefont {Yuan},
  \citenamefont {Watanabe}, \citenamefont {Taniguchi}, \citenamefont
  {Fernandes}, \citenamefont {Fu},\ and\ \citenamefont
  {Jarillo-Herrero}}]{cao_nematicity_2021}%
  \BibitemOpen
  \bibfield  {author} {\bibinfo {author} {\bibfnamefont {Y.}~\bibnamefont
  {Cao}}, \bibinfo {author} {\bibfnamefont {D.}~\bibnamefont {Rodan-Legrain}},
  \bibinfo {author} {\bibfnamefont {J.~M.}\ \bibnamefont {Park}}, \bibinfo
  {author} {\bibfnamefont {N.~F.~Q.}\ \bibnamefont {Yuan}}, \bibinfo {author}
  {\bibfnamefont {K.}~\bibnamefont {Watanabe}}, \bibinfo {author}
  {\bibfnamefont {T.}~\bibnamefont {Taniguchi}}, \bibinfo {author}
  {\bibfnamefont {R.~M.}\ \bibnamefont {Fernandes}}, \bibinfo {author}
  {\bibfnamefont {L.}~\bibnamefont {Fu}}, \ and\ \bibinfo {author}
  {\bibfnamefont {P.}~\bibnamefont {Jarillo-Herrero}},\ }\href {\doibase
  10.1126/science.abc2836} {\bibfield  {journal} {\bibinfo  {journal}
  {Science}\ }\textbf {\bibinfo {volume} {372}},\ \bibinfo {pages} {264}
  (\bibinfo {year} {2021})}\BibitemShut {NoStop}%
\bibitem [{\citenamefont {Khalaf}\ \emph {et~al.}(2021)\citenamefont {Khalaf},
  \citenamefont {Chatterjee}, \citenamefont {Bultinck}, \citenamefont
  {Zaletel},\ and\ \citenamefont {Vishwanath}}]{khalaf_charged_2021}%
  \BibitemOpen
  \bibfield  {author} {\bibinfo {author} {\bibfnamefont {E.}~\bibnamefont
  {Khalaf}}, \bibinfo {author} {\bibfnamefont {S.}~\bibnamefont {Chatterjee}},
  \bibinfo {author} {\bibfnamefont {N.}~\bibnamefont {Bultinck}}, \bibinfo
  {author} {\bibfnamefont {M.~P.}\ \bibnamefont {Zaletel}}, \ and\ \bibinfo
  {author} {\bibfnamefont {A.}~\bibnamefont {Vishwanath}},\ }\href {\doibase
  10.1126/sciadv.abf5299} {\bibfield  {journal} {\bibinfo  {journal} {Science
  Advances}\ }\textbf {\bibinfo {volume} {7}},\ \bibinfo {pages} {eabf5299}
  (\bibinfo {year} {2021})}\BibitemShut {NoStop}%
\bibitem [{\citenamefont {Chew}\ \emph {et~al.}(2023)\citenamefont {Chew},
  \citenamefont {Wang}, \citenamefont {Bernevig},\ and\ \citenamefont
  {Song}}]{chew_higher-order_2023}%
  \BibitemOpen
  \bibfield  {author} {\bibinfo {author} {\bibfnamefont {A.}~\bibnamefont
  {Chew}}, \bibinfo {author} {\bibfnamefont {Y.}~\bibnamefont {Wang}}, \bibinfo
  {author} {\bibfnamefont {B.~A.}\ \bibnamefont {Bernevig}}, \ and\ \bibinfo
  {author} {\bibfnamefont {Z.-D.}\ \bibnamefont {Song}},\ }\href {\doibase
  10.1103/PhysRevB.107.094512} {\bibfield  {journal} {\bibinfo  {journal}
  {Physical Review B}\ }\textbf {\bibinfo {volume} {107}},\ \bibinfo {pages}
  {094512} (\bibinfo {year} {2023})}\BibitemShut {NoStop}%
\bibitem [{\citenamefont {Stepanov}\ \emph {et~al.}(2021)\citenamefont
  {Stepanov}, \citenamefont {Xie}, \citenamefont {Taniguchi}, \citenamefont
  {Watanabe}, \citenamefont {Lu}, \citenamefont {MacDonald}, \citenamefont
  {Bernevig},\ and\ \citenamefont {Efetov}}]{stepanov_competing_2021}%
  \BibitemOpen
  \bibfield  {author} {\bibinfo {author} {\bibfnamefont {P.}~\bibnamefont
  {Stepanov}}, \bibinfo {author} {\bibfnamefont {M.}~\bibnamefont {Xie}},
  \bibinfo {author} {\bibfnamefont {T.}~\bibnamefont {Taniguchi}}, \bibinfo
  {author} {\bibfnamefont {K.}~\bibnamefont {Watanabe}}, \bibinfo {author}
  {\bibfnamefont {X.}~\bibnamefont {Lu}}, \bibinfo {author} {\bibfnamefont
  {A.}~\bibnamefont {MacDonald}}, \bibinfo {author} {\bibfnamefont {B.~A.}\
  \bibnamefont {Bernevig}}, \ and\ \bibinfo {author} {\bibfnamefont
  {D.}~\bibnamefont {Efetov}},\ }\href {\doibase
  10.1103/PhysRevLett.127.197701} {\bibfield  {journal} {\bibinfo  {journal}
  {Physical Review Letters}\ }\textbf {\bibinfo {volume} {127}},\ \bibinfo
  {pages} {197701} (\bibinfo {year} {2021})}\BibitemShut {NoStop}%
\bibitem [{\citenamefont {Banszerus}\ \emph
  {et~al.}(2020{\natexlab{a}})\citenamefont {Banszerus}, \citenamefont
  {Rothstein}, \citenamefont {Fabian}, \citenamefont {Möller}, \citenamefont
  {Icking}, \citenamefont {Trellenkamp}, \citenamefont {Lentz}, \citenamefont
  {Neumaier}, \citenamefont {Watanabe}, \citenamefont {Taniguchi},
  \citenamefont {Libisch}, \citenamefont {Volk},\ and\ \citenamefont
  {Stampfer}}]{banszerus_electronhole_2020}%
  \BibitemOpen
  \bibfield  {author} {\bibinfo {author} {\bibfnamefont {L.}~\bibnamefont
  {Banszerus}}, \bibinfo {author} {\bibfnamefont {A.}~\bibnamefont
  {Rothstein}}, \bibinfo {author} {\bibfnamefont {T.}~\bibnamefont {Fabian}},
  \bibinfo {author} {\bibfnamefont {S.}~\bibnamefont {Möller}}, \bibinfo
  {author} {\bibfnamefont {E.}~\bibnamefont {Icking}}, \bibinfo {author}
  {\bibfnamefont {S.}~\bibnamefont {Trellenkamp}}, \bibinfo {author}
  {\bibfnamefont {F.}~\bibnamefont {Lentz}}, \bibinfo {author} {\bibfnamefont
  {D.}~\bibnamefont {Neumaier}}, \bibinfo {author} {\bibfnamefont
  {K.}~\bibnamefont {Watanabe}}, \bibinfo {author} {\bibfnamefont
  {T.}~\bibnamefont {Taniguchi}}, \bibinfo {author} {\bibfnamefont
  {F.}~\bibnamefont {Libisch}}, \bibinfo {author} {\bibfnamefont
  {C.}~\bibnamefont {Volk}}, \ and\ \bibinfo {author} {\bibfnamefont
  {C.}~\bibnamefont {Stampfer}},\ }\href {\doibase
  10.1021/acs.nanolett.0c03227} {\bibfield  {journal} {\bibinfo  {journal}
  {Nano Letters}\ }\textbf {\bibinfo {volume} {20}},\ \bibinfo {pages} {7709}
  (\bibinfo {year} {2020}{\natexlab{a}})}\BibitemShut {NoStop}%
\bibitem [{\citenamefont {Banszerus}\ \emph
  {et~al.}(2020{\natexlab{b}})\citenamefont {Banszerus}, \citenamefont
  {Möller}, \citenamefont {Icking}, \citenamefont {Watanabe}, \citenamefont
  {Taniguchi}, \citenamefont {Volk},\ and\ \citenamefont
  {Stampfer}}]{banszerus_single-electron_2020}%
  \BibitemOpen
  \bibfield  {author} {\bibinfo {author} {\bibfnamefont {L.}~\bibnamefont
  {Banszerus}}, \bibinfo {author} {\bibfnamefont {S.}~\bibnamefont {Möller}},
  \bibinfo {author} {\bibfnamefont {E.}~\bibnamefont {Icking}}, \bibinfo
  {author} {\bibfnamefont {K.}~\bibnamefont {Watanabe}}, \bibinfo {author}
  {\bibfnamefont {T.}~\bibnamefont {Taniguchi}}, \bibinfo {author}
  {\bibfnamefont {C.}~\bibnamefont {Volk}}, \ and\ \bibinfo {author}
  {\bibfnamefont {C.}~\bibnamefont {Stampfer}},\ }\href {\doibase
  10.1021/acs.nanolett.9b05295} {\bibfield  {journal} {\bibinfo  {journal}
  {Nano Letters}\ }\textbf {\bibinfo {volume} {20}},\ \bibinfo {pages} {2005}
  (\bibinfo {year} {2020}{\natexlab{b}})}\BibitemShut {NoStop}%
\bibitem [{\citenamefont {Ge}\ \emph {et~al.}(2020)\citenamefont {Ge},
  \citenamefont {Joucken}, \citenamefont {Quezada}, \citenamefont {Da~Costa},
  \citenamefont {Davenport}, \citenamefont {Giraldo}, \citenamefont
  {Taniguchi}, \citenamefont {Watanabe}, \citenamefont {Kobayashi},
  \citenamefont {Low},\ and\ \citenamefont {Velasco}}]{ge_visualization_2020}%
  \BibitemOpen
  \bibfield  {author} {\bibinfo {author} {\bibfnamefont {Z.}~\bibnamefont
  {Ge}}, \bibinfo {author} {\bibfnamefont {F.}~\bibnamefont {Joucken}},
  \bibinfo {author} {\bibfnamefont {E.}~\bibnamefont {Quezada}}, \bibinfo
  {author} {\bibfnamefont {D.~R.}\ \bibnamefont {Da~Costa}}, \bibinfo {author}
  {\bibfnamefont {J.}~\bibnamefont {Davenport}}, \bibinfo {author}
  {\bibfnamefont {B.}~\bibnamefont {Giraldo}}, \bibinfo {author} {\bibfnamefont
  {T.}~\bibnamefont {Taniguchi}}, \bibinfo {author} {\bibfnamefont
  {K.}~\bibnamefont {Watanabe}}, \bibinfo {author} {\bibfnamefont {N.~P.}\
  \bibnamefont {Kobayashi}}, \bibinfo {author} {\bibfnamefont {T.}~\bibnamefont
  {Low}}, \ and\ \bibinfo {author} {\bibfnamefont {J.}~\bibnamefont
  {Velasco}},\ }\href {\doibase 10.1021/acs.nanolett.0c03453} {\bibfield
  {journal} {\bibinfo  {journal} {Nano Letters}\ }\textbf {\bibinfo {volume}
  {20}},\ \bibinfo {pages} {8682} (\bibinfo {year} {2020})}\BibitemShut
  {NoStop}%
\bibitem [{\citenamefont {Banszerus}\ \emph {et~al.}(2021)\citenamefont
  {Banszerus}, \citenamefont {Möller}, \citenamefont {Steiner}, \citenamefont
  {Icking}, \citenamefont {Trellenkamp}, \citenamefont {Lentz}, \citenamefont
  {Watanabe}, \citenamefont {Taniguchi}, \citenamefont {Volk},\ and\
  \citenamefont {Stampfer}}]{banszerus_spin-valley_2021}%
  \BibitemOpen
  \bibfield  {author} {\bibinfo {author} {\bibfnamefont {L.}~\bibnamefont
  {Banszerus}}, \bibinfo {author} {\bibfnamefont {S.}~\bibnamefont {Möller}},
  \bibinfo {author} {\bibfnamefont {C.}~\bibnamefont {Steiner}}, \bibinfo
  {author} {\bibfnamefont {E.}~\bibnamefont {Icking}}, \bibinfo {author}
  {\bibfnamefont {S.}~\bibnamefont {Trellenkamp}}, \bibinfo {author}
  {\bibfnamefont {F.}~\bibnamefont {Lentz}}, \bibinfo {author} {\bibfnamefont
  {K.}~\bibnamefont {Watanabe}}, \bibinfo {author} {\bibfnamefont
  {T.}~\bibnamefont {Taniguchi}}, \bibinfo {author} {\bibfnamefont
  {C.}~\bibnamefont {Volk}}, \ and\ \bibinfo {author} {\bibfnamefont
  {C.}~\bibnamefont {Stampfer}},\ }\href {\doibase 10.1038/s41467-021-25498-3}
  {\bibfield  {journal} {\bibinfo  {journal} {Nature Communications}\ }\textbf
  {\bibinfo {volume} {12}},\ \bibinfo {pages} {5250} (\bibinfo {year}
  {2021})}\BibitemShut {NoStop}%
\bibitem [{\citenamefont {Solomon}\ and\ \citenamefont
  {Power}(2021)}]{solomon_valley_2021}%
  \BibitemOpen
  \bibfield  {author} {\bibinfo {author} {\bibfnamefont {F.}~\bibnamefont
  {Solomon}}\ and\ \bibinfo {author} {\bibfnamefont {S.~R.}\ \bibnamefont
  {Power}},\ }\href {\doibase 10.1103/PhysRevB.103.235435} {\bibfield
  {journal} {\bibinfo  {journal} {Physical Review B}\ }\textbf {\bibinfo
  {volume} {103}},\ \bibinfo {pages} {235435} (\bibinfo {year}
  {2021})}\BibitemShut {NoStop}%
\bibitem [{\citenamefont {Hecker}\ \emph {et~al.}(2023)\citenamefont {Hecker},
  \citenamefont {Banszerus}, \citenamefont {Schäpers}, \citenamefont
  {Möller}, \citenamefont {Peters}, \citenamefont {Icking}, \citenamefont
  {Watanabe}, \citenamefont {Taniguchi}, \citenamefont {Volk},\ and\
  \citenamefont {Stampfer}}]{hecker_coherent_2023}%
  \BibitemOpen
  \bibfield  {author} {\bibinfo {author} {\bibfnamefont {K.}~\bibnamefont
  {Hecker}}, \bibinfo {author} {\bibfnamefont {L.}~\bibnamefont {Banszerus}},
  \bibinfo {author} {\bibfnamefont {A.}~\bibnamefont {Schäpers}}, \bibinfo
  {author} {\bibfnamefont {S.}~\bibnamefont {Möller}}, \bibinfo {author}
  {\bibfnamefont {A.}~\bibnamefont {Peters}}, \bibinfo {author} {\bibfnamefont
  {E.}~\bibnamefont {Icking}}, \bibinfo {author} {\bibfnamefont
  {K.}~\bibnamefont {Watanabe}}, \bibinfo {author} {\bibfnamefont
  {T.}~\bibnamefont {Taniguchi}}, \bibinfo {author} {\bibfnamefont
  {C.}~\bibnamefont {Volk}}, \ and\ \bibinfo {author} {\bibfnamefont
  {C.}~\bibnamefont {Stampfer}},\ }\href {\doibase 10.1038/s41467-023-43541-3}
  {\bibfield  {journal} {\bibinfo  {journal} {Nature Communications}\ }\textbf
  {\bibinfo {volume} {14}},\ \bibinfo {pages} {7911} (\bibinfo {year}
  {2023})}\BibitemShut {NoStop}%
\bibitem [{\citenamefont {Korkusinski}\ \emph {et~al.}(2023)\citenamefont
  {Korkusinski}, \citenamefont {Saleem}, \citenamefont {Dusko}, \citenamefont
  {Miravet},\ and\ \citenamefont {Hawrylak}}]{korkusinski_spontaneous_2023}%
  \BibitemOpen
  \bibfield  {author} {\bibinfo {author} {\bibfnamefont {M.}~\bibnamefont
  {Korkusinski}}, \bibinfo {author} {\bibfnamefont {Y.}~\bibnamefont {Saleem}},
  \bibinfo {author} {\bibfnamefont {A.}~\bibnamefont {Dusko}}, \bibinfo
  {author} {\bibfnamefont {D.}~\bibnamefont {Miravet}}, \ and\ \bibinfo
  {author} {\bibfnamefont {P.}~\bibnamefont {Hawrylak}},\ }\href {\doibase
  10.1021/acs.nanolett.3c02073} {\bibfield  {journal} {\bibinfo  {journal}
  {Nano Letters}\ }\textbf {\bibinfo {volume} {23}},\ \bibinfo {pages} {7546}
  (\bibinfo {year} {2023})}\BibitemShut {NoStop}%
\bibitem [{\citenamefont {Garreis}\ \emph {et~al.}(2024)\citenamefont
  {Garreis}, \citenamefont {Tong}, \citenamefont {Terle}, \citenamefont
  {Ruckriegel}, \citenamefont {Gerber}, \citenamefont {Gächter}, \citenamefont
  {Watanabe}, \citenamefont {Taniguchi}, \citenamefont {Ihn}, \citenamefont
  {Ensslin},\ and\ \citenamefont {Huang}}]{garreis_long-lived_2024}%
  \BibitemOpen
  \bibfield  {author} {\bibinfo {author} {\bibfnamefont {R.}~\bibnamefont
  {Garreis}}, \bibinfo {author} {\bibfnamefont {C.}~\bibnamefont {Tong}},
  \bibinfo {author} {\bibfnamefont {J.}~\bibnamefont {Terle}}, \bibinfo
  {author} {\bibfnamefont {M.~J.}\ \bibnamefont {Ruckriegel}}, \bibinfo
  {author} {\bibfnamefont {J.~D.}\ \bibnamefont {Gerber}}, \bibinfo {author}
  {\bibfnamefont {L.~M.}\ \bibnamefont {Gächter}}, \bibinfo {author}
  {\bibfnamefont {K.}~\bibnamefont {Watanabe}}, \bibinfo {author}
  {\bibfnamefont {T.}~\bibnamefont {Taniguchi}}, \bibinfo {author}
  {\bibfnamefont {T.}~\bibnamefont {Ihn}}, \bibinfo {author} {\bibfnamefont
  {K.}~\bibnamefont {Ensslin}}, \ and\ \bibinfo {author} {\bibfnamefont
  {W.~W.}\ \bibnamefont {Huang}},\ }\href {\doibase 10.1038/s41567-023-02334-7}
  {\bibfield  {journal} {\bibinfo  {journal} {Nature Physics}\ }\textbf
  {\bibinfo {volume} {20}},\ \bibinfo {pages} {428} (\bibinfo {year}
  {2024})}\BibitemShut {NoStop}%
\bibitem [{\citenamefont {McCann}(2006)}]{mccann_asymmetry_2006}%
  \BibitemOpen
  \bibfield  {author} {\bibinfo {author} {\bibfnamefont {E.}~\bibnamefont
  {McCann}},\ }\href {\doibase 10.1103/PhysRevB.74.161403} {\bibfield
  {journal} {\bibinfo  {journal} {Physical Review B}\ }\textbf {\bibinfo
  {volume} {74}},\ \bibinfo {pages} {161403} (\bibinfo {year}
  {2006})}\BibitemShut {NoStop}%
\bibitem [{\citenamefont {Fogler}\ and\ \citenamefont
  {McCann}(2010)}]{fogler_comment_2010}%
  \BibitemOpen
  \bibfield  {author} {\bibinfo {author} {\bibfnamefont {M.~M.}\ \bibnamefont
  {Fogler}}\ and\ \bibinfo {author} {\bibfnamefont {E.}~\bibnamefont
  {McCann}},\ }\href {\doibase 10.1103/PhysRevB.82.197401} {\bibfield
  {journal} {\bibinfo  {journal} {Physical Review B}\ }\textbf {\bibinfo
  {volume} {82}},\ \bibinfo {pages} {197401} (\bibinfo {year}
  {2010})}\BibitemShut {NoStop}%
\bibitem [{\citenamefont {Chernikov}\ \emph {et~al.}(2015)\citenamefont
  {Chernikov}, \citenamefont {Ruppert}, \citenamefont {Hill}, \citenamefont
  {Rigosi},\ and\ \citenamefont {Heinz}}]{chernikov_population_2015}%
  \BibitemOpen
  \bibfield  {author} {\bibinfo {author} {\bibfnamefont {A.}~\bibnamefont
  {Chernikov}}, \bibinfo {author} {\bibfnamefont {C.}~\bibnamefont {Ruppert}},
  \bibinfo {author} {\bibfnamefont {H.~M.}\ \bibnamefont {Hill}}, \bibinfo
  {author} {\bibfnamefont {A.~F.}\ \bibnamefont {Rigosi}}, \ and\ \bibinfo
  {author} {\bibfnamefont {T.~F.}\ \bibnamefont {Heinz}},\ }\href {\doibase
  10.1038/nphoton.2015.104} {\bibfield  {journal} {\bibinfo  {journal} {Nature
  Photonics}\ }\textbf {\bibinfo {volume} {9}},\ \bibinfo {pages} {466}
  (\bibinfo {year} {2015})}\BibitemShut {NoStop}%
\bibitem [{\citenamefont {Pogna}\ \emph {et~al.}(2016)\citenamefont {Pogna},
  \citenamefont {Marsili}, \citenamefont {De~Fazio}, \citenamefont {Dal~Conte},
  \citenamefont {Manzoni}, \citenamefont {Sangalli}, \citenamefont {Yoon},
  \citenamefont {Lombardo}, \citenamefont {Ferrari}, \citenamefont {Marini},
  \citenamefont {Cerullo},\ and\ \citenamefont
  {Prezzi}}]{pogna_photo-induced_2016}%
  \BibitemOpen
  \bibfield  {author} {\bibinfo {author} {\bibfnamefont {E.~A.~A.}\
  \bibnamefont {Pogna}}, \bibinfo {author} {\bibfnamefont {M.}~\bibnamefont
  {Marsili}}, \bibinfo {author} {\bibfnamefont {D.}~\bibnamefont {De~Fazio}},
  \bibinfo {author} {\bibfnamefont {S.}~\bibnamefont {Dal~Conte}}, \bibinfo
  {author} {\bibfnamefont {C.}~\bibnamefont {Manzoni}}, \bibinfo {author}
  {\bibfnamefont {D.}~\bibnamefont {Sangalli}}, \bibinfo {author}
  {\bibfnamefont {D.}~\bibnamefont {Yoon}}, \bibinfo {author} {\bibfnamefont
  {A.}~\bibnamefont {Lombardo}}, \bibinfo {author} {\bibfnamefont {A.~C.}\
  \bibnamefont {Ferrari}}, \bibinfo {author} {\bibfnamefont {A.}~\bibnamefont
  {Marini}}, \bibinfo {author} {\bibfnamefont {G.}~\bibnamefont {Cerullo}}, \
  and\ \bibinfo {author} {\bibfnamefont {D.}~\bibnamefont {Prezzi}},\ }\href
  {\doibase 10.1021/acsnano.5b06488} {\bibfield  {journal} {\bibinfo  {journal}
  {ACS Nano}\ }\textbf {\bibinfo {volume} {10}},\ \bibinfo {pages} {1182}
  (\bibinfo {year} {2016})}\BibitemShut {NoStop}%
\bibitem [{\citenamefont {Cunningham}\ \emph {et~al.}(2017)\citenamefont
  {Cunningham}, \citenamefont {Hanbicki}, \citenamefont {McCreary},\ and\
  \citenamefont {Jonker}}]{cunningham_photoinduced_2017}%
  \BibitemOpen
  \bibfield  {author} {\bibinfo {author} {\bibfnamefont {P.~D.}\ \bibnamefont
  {Cunningham}}, \bibinfo {author} {\bibfnamefont {A.~T.}\ \bibnamefont
  {Hanbicki}}, \bibinfo {author} {\bibfnamefont {K.~M.}\ \bibnamefont
  {McCreary}}, \ and\ \bibinfo {author} {\bibfnamefont {B.~T.}\ \bibnamefont
  {Jonker}},\ }\href {\doibase 10.1021/acsnano.7b06885} {\bibfield  {journal}
  {\bibinfo  {journal} {ACS Nano}\ }\textbf {\bibinfo {volume} {11}},\ \bibinfo
  {pages} {12601} (\bibinfo {year} {2017})}\BibitemShut {NoStop}%
\bibitem [{\citenamefont {Yao}\ \emph {et~al.}(2017)\citenamefont {Yao},
  \citenamefont {Yan}, \citenamefont {Kahn}, \citenamefont {Suslu},
  \citenamefont {Liang}, \citenamefont {Barnard}, \citenamefont {Tongay},
  \citenamefont {Zettl}, \citenamefont {Borys},\ and\ \citenamefont
  {Schuck}}]{yao_optically_2017}%
  \BibitemOpen
  \bibfield  {author} {\bibinfo {author} {\bibfnamefont {K.}~\bibnamefont
  {Yao}}, \bibinfo {author} {\bibfnamefont {A.}~\bibnamefont {Yan}}, \bibinfo
  {author} {\bibfnamefont {S.}~\bibnamefont {Kahn}}, \bibinfo {author}
  {\bibfnamefont {A.}~\bibnamefont {Suslu}}, \bibinfo {author} {\bibfnamefont
  {Y.}~\bibnamefont {Liang}}, \bibinfo {author} {\bibfnamefont {E.~S.}\
  \bibnamefont {Barnard}}, \bibinfo {author} {\bibfnamefont {S.}~\bibnamefont
  {Tongay}}, \bibinfo {author} {\bibfnamefont {A.}~\bibnamefont {Zettl}},
  \bibinfo {author} {\bibfnamefont {N.}~\bibnamefont {Borys}}, \ and\ \bibinfo
  {author} {\bibfnamefont {P.~J.}\ \bibnamefont {Schuck}},\ }\href {\doibase
  10.1103/PhysRevLett.119.087401} {\bibfield  {journal} {\bibinfo  {journal}
  {Physical Review Letters}\ }\textbf {\bibinfo {volume} {119}},\ \bibinfo
  {pages} {087401} (\bibinfo {year} {2017})}\BibitemShut {NoStop}%
\bibitem [{\citenamefont {Qiu}\ \emph {et~al.}(2019)\citenamefont {Qiu},
  \citenamefont {Trushin}, \citenamefont {Fang}, \citenamefont {Verzhbitskiy},
  \citenamefont {Gao}, \citenamefont {Laksono}, \citenamefont {Yang},
  \citenamefont {Lyu}, \citenamefont {Li}, \citenamefont {Su}, \citenamefont
  {Telychko}, \citenamefont {Watanabe}, \citenamefont {Taniguchi},
  \citenamefont {Wu}, \citenamefont {Neto}, \citenamefont {Yang}, \citenamefont
  {Eda}, \citenamefont {Adam},\ and\ \citenamefont {Lu}}]{qiu_giant_2019}%
  \BibitemOpen
  \bibfield  {author} {\bibinfo {author} {\bibfnamefont {Z.}~\bibnamefont
  {Qiu}}, \bibinfo {author} {\bibfnamefont {M.}~\bibnamefont {Trushin}},
  \bibinfo {author} {\bibfnamefont {H.}~\bibnamefont {Fang}}, \bibinfo {author}
  {\bibfnamefont {I.}~\bibnamefont {Verzhbitskiy}}, \bibinfo {author}
  {\bibfnamefont {S.}~\bibnamefont {Gao}}, \bibinfo {author} {\bibfnamefont
  {E.}~\bibnamefont {Laksono}}, \bibinfo {author} {\bibfnamefont
  {M.}~\bibnamefont {Yang}}, \bibinfo {author} {\bibfnamefont {P.}~\bibnamefont
  {Lyu}}, \bibinfo {author} {\bibfnamefont {J.}~\bibnamefont {Li}}, \bibinfo
  {author} {\bibfnamefont {J.}~\bibnamefont {Su}}, \bibinfo {author}
  {\bibfnamefont {M.}~\bibnamefont {Telychko}}, \bibinfo {author}
  {\bibfnamefont {K.}~\bibnamefont {Watanabe}}, \bibinfo {author}
  {\bibfnamefont {T.}~\bibnamefont {Taniguchi}}, \bibinfo {author}
  {\bibfnamefont {J.}~\bibnamefont {Wu}}, \bibinfo {author} {\bibfnamefont
  {A.~H.~C.}\ \bibnamefont {Neto}}, \bibinfo {author} {\bibfnamefont
  {L.}~\bibnamefont {Yang}}, \bibinfo {author} {\bibfnamefont {G.}~\bibnamefont
  {Eda}}, \bibinfo {author} {\bibfnamefont {S.}~\bibnamefont {Adam}}, \ and\
  \bibinfo {author} {\bibfnamefont {J.}~\bibnamefont {Lu}},\ }\href {\doibase
  10.1126/sciadv.aaw2347} {\bibfield  {journal} {\bibinfo  {journal} {Science
  Advances}\ }\textbf {\bibinfo {volume} {5}},\ \bibinfo {pages} {eaaw2347}
  (\bibinfo {year} {2019})}\BibitemShut {NoStop}%
\bibitem [{\citenamefont {Bera}\ \emph {et~al.}(2021)\citenamefont {Bera},
  \citenamefont {Shrivastava}, \citenamefont {Bramhachari}, \citenamefont
  {Zhang}, \citenamefont {Poonia}, \citenamefont {Mandal}, \citenamefont
  {Miller}, \citenamefont {Beard}, \citenamefont {Agarwal},\ and\ \citenamefont
  {Adarsh}}]{bera_atomlike_2021}%
  \BibitemOpen
  \bibfield  {author} {\bibinfo {author} {\bibfnamefont {S.~K.}\ \bibnamefont
  {Bera}}, \bibinfo {author} {\bibfnamefont {M.}~\bibnamefont {Shrivastava}},
  \bibinfo {author} {\bibfnamefont {K.}~\bibnamefont {Bramhachari}}, \bibinfo
  {author} {\bibfnamefont {H.}~\bibnamefont {Zhang}}, \bibinfo {author}
  {\bibfnamefont {A.~K.}\ \bibnamefont {Poonia}}, \bibinfo {author}
  {\bibfnamefont {D.}~\bibnamefont {Mandal}}, \bibinfo {author} {\bibfnamefont
  {E.~M.}\ \bibnamefont {Miller}}, \bibinfo {author} {\bibfnamefont {M.~C.}\
  \bibnamefont {Beard}}, \bibinfo {author} {\bibfnamefont {A.}~\bibnamefont
  {Agarwal}}, \ and\ \bibinfo {author} {\bibfnamefont {K.~V.}\ \bibnamefont
  {Adarsh}},\ }\href {\doibase 10.1103/PhysRevB.104.L201404} {\bibfield
  {journal} {\bibinfo  {journal} {Physical Review B}\ }\textbf {\bibinfo
  {volume} {104}},\ \bibinfo {pages} {L201404} (\bibinfo {year}
  {2021})}\BibitemShut {NoStop}%
\bibitem [{\citenamefont {Kang}\ \emph {et~al.}(2017)\citenamefont {Kang},
  \citenamefont {Kim}, \citenamefont {Ryu}, \citenamefont {Jung}, \citenamefont
  {Kim}, \citenamefont {Moreschini}, \citenamefont {Jozwiak}, \citenamefont
  {Rotenberg}, \citenamefont {Bostwick},\ and\ \citenamefont
  {Kim}}]{kang_universal_2017}%
  \BibitemOpen
  \bibfield  {author} {\bibinfo {author} {\bibfnamefont {M.}~\bibnamefont
  {Kang}}, \bibinfo {author} {\bibfnamefont {B.}~\bibnamefont {Kim}}, \bibinfo
  {author} {\bibfnamefont {S.~H.}\ \bibnamefont {Ryu}}, \bibinfo {author}
  {\bibfnamefont {S.~W.}\ \bibnamefont {Jung}}, \bibinfo {author}
  {\bibfnamefont {J.}~\bibnamefont {Kim}}, \bibinfo {author} {\bibfnamefont
  {L.}~\bibnamefont {Moreschini}}, \bibinfo {author} {\bibfnamefont
  {C.}~\bibnamefont {Jozwiak}}, \bibinfo {author} {\bibfnamefont
  {E.}~\bibnamefont {Rotenberg}}, \bibinfo {author} {\bibfnamefont
  {A.}~\bibnamefont {Bostwick}}, \ and\ \bibinfo {author} {\bibfnamefont
  {K.~S.}\ \bibnamefont {Kim}},\ }\href {\doibase 10.1021/acs.nanolett.6b04775}
  {\bibfield  {journal} {\bibinfo  {journal} {Nano Letters}\ }\textbf {\bibinfo
  {volume} {17}},\ \bibinfo {pages} {1610} (\bibinfo {year}
  {2017})}\BibitemShut {NoStop}%
\bibitem [{\citenamefont {Nguyen}\ \emph {et~al.}(2019)\citenamefont {Nguyen},
  \citenamefont {Teutsch}, \citenamefont {Wilson}, \citenamefont {Kahn},
  \citenamefont {Xia}, \citenamefont {Graham}, \citenamefont {Kandyba},
  \citenamefont {Giampietri}, \citenamefont {Barinov}, \citenamefont
  {Constantinescu}, \citenamefont {Yeung}, \citenamefont {Hine}, \citenamefont
  {Xu}, \citenamefont {Cobden},\ and\ \citenamefont
  {Wilson}}]{nguyen_visualizing_2019}%
  \BibitemOpen
  \bibfield  {author} {\bibinfo {author} {\bibfnamefont {P.~V.}\ \bibnamefont
  {Nguyen}}, \bibinfo {author} {\bibfnamefont {N.~C.}\ \bibnamefont {Teutsch}},
  \bibinfo {author} {\bibfnamefont {N.~P.}\ \bibnamefont {Wilson}}, \bibinfo
  {author} {\bibfnamefont {J.}~\bibnamefont {Kahn}}, \bibinfo {author}
  {\bibfnamefont {X.}~\bibnamefont {Xia}}, \bibinfo {author} {\bibfnamefont
  {A.~J.}\ \bibnamefont {Graham}}, \bibinfo {author} {\bibfnamefont
  {V.}~\bibnamefont {Kandyba}}, \bibinfo {author} {\bibfnamefont
  {A.}~\bibnamefont {Giampietri}}, \bibinfo {author} {\bibfnamefont
  {A.}~\bibnamefont {Barinov}}, \bibinfo {author} {\bibfnamefont {G.~C.}\
  \bibnamefont {Constantinescu}}, \bibinfo {author} {\bibfnamefont
  {N.}~\bibnamefont {Yeung}}, \bibinfo {author} {\bibfnamefont {N.~D.~M.}\
  \bibnamefont {Hine}}, \bibinfo {author} {\bibfnamefont {X.}~\bibnamefont
  {Xu}}, \bibinfo {author} {\bibfnamefont {D.~H.}\ \bibnamefont {Cobden}}, \
  and\ \bibinfo {author} {\bibfnamefont {N.~R.}\ \bibnamefont {Wilson}},\
  }\href {\doibase 10.1038/s41586-019-1402-1} {\bibfield  {journal} {\bibinfo
  {journal} {Nature}\ }\textbf {\bibinfo {volume} {572}},\ \bibinfo {pages}
  {220} (\bibinfo {year} {2019})}\BibitemShut {NoStop}%
\bibitem [{\citenamefont {Liu}\ \emph {et~al.}(2019)\citenamefont {Liu},
  \citenamefont {Ziffer}, \citenamefont {Hansen}, \citenamefont {Wang},\ and\
  \citenamefont {Zhu}}]{liu_direct_2019}%
  \BibitemOpen
  \bibfield  {author} {\bibinfo {author} {\bibfnamefont {F.}~\bibnamefont
  {Liu}}, \bibinfo {author} {\bibfnamefont {M.~E.}\ \bibnamefont {Ziffer}},
  \bibinfo {author} {\bibfnamefont {K.~R.}\ \bibnamefont {Hansen}}, \bibinfo
  {author} {\bibfnamefont {J.}~\bibnamefont {Wang}}, \ and\ \bibinfo {author}
  {\bibfnamefont {X.}~\bibnamefont {Zhu}},\ }\href {\doibase
  10.1103/PhysRevLett.122.246803} {\bibfield  {journal} {\bibinfo  {journal}
  {Physical Review Letters}\ }\textbf {\bibinfo {volume} {122}},\ \bibinfo
  {pages} {246803} (\bibinfo {year} {2019})}\BibitemShut {NoStop}%
\bibitem [{\citenamefont {Gao}\ \emph {et~al.}(2016)\citenamefont {Gao},
  \citenamefont {Liang}, \citenamefont {Spataru},\ and\ \citenamefont
  {Yang}}]{gao_dynamical_2016}%
  \BibitemOpen
  \bibfield  {author} {\bibinfo {author} {\bibfnamefont {S.}~\bibnamefont
  {Gao}}, \bibinfo {author} {\bibfnamefont {Y.}~\bibnamefont {Liang}}, \bibinfo
  {author} {\bibfnamefont {C.~D.}\ \bibnamefont {Spataru}}, \ and\ \bibinfo
  {author} {\bibfnamefont {L.}~\bibnamefont {Yang}},\ }\href {\doibase
  10.1021/acs.nanolett.6b02118} {\bibfield  {journal} {\bibinfo  {journal}
  {Nano Letters}\ }\textbf {\bibinfo {volume} {16}},\ \bibinfo {pages} {5568}
  (\bibinfo {year} {2016})}\BibitemShut {NoStop}%
\bibitem [{\citenamefont {Gao}\ and\ \citenamefont
  {Yang}(2017)}]{gao_renormalization_2017}%
  \BibitemOpen
  \bibfield  {author} {\bibinfo {author} {\bibfnamefont {S.}~\bibnamefont
  {Gao}}\ and\ \bibinfo {author} {\bibfnamefont {L.}~\bibnamefont {Yang}},\
  }\href {\doibase 10.1103/PhysRevB.96.155410} {\bibfield  {journal} {\bibinfo
  {journal} {Physical Review B}\ }\textbf {\bibinfo {volume} {96}},\ \bibinfo
  {pages} {155410} (\bibinfo {year} {2017})}\BibitemShut {NoStop}%
\bibitem [{\citenamefont {Liang}\ and\ \citenamefont
  {Yang}(2015)}]{liang_carrier_2015}%
  \BibitemOpen
  \bibfield  {author} {\bibinfo {author} {\bibfnamefont {Y.}~\bibnamefont
  {Liang}}\ and\ \bibinfo {author} {\bibfnamefont {L.}~\bibnamefont {Yang}},\
  }\href {\doibase 10.1103/PhysRevLett.114.063001} {\bibfield  {journal}
  {\bibinfo  {journal} {Physical Review Letters}\ }\textbf {\bibinfo {volume}
  {114}},\ \bibinfo {pages} {063001} (\bibinfo {year} {2015})}\BibitemShut
  {NoStop}%
\bibitem [{\citenamefont {Faridi}\ \emph {et~al.}(2021)\citenamefont {Faridi},
  \citenamefont {Culcer},\ and\ \citenamefont
  {Asgari}}]{faridi_quasiparticle_2021}%
  \BibitemOpen
  \bibfield  {author} {\bibinfo {author} {\bibfnamefont {A.}~\bibnamefont
  {Faridi}}, \bibinfo {author} {\bibfnamefont {D.}~\bibnamefont {Culcer}}, \
  and\ \bibinfo {author} {\bibfnamefont {R.}~\bibnamefont {Asgari}},\ }\href
  {\doibase 10.1103/PhysRevB.104.085432} {\bibfield  {journal} {\bibinfo
  {journal} {Physical Review B}\ }\textbf {\bibinfo {volume} {104}},\ \bibinfo
  {pages} {085432} (\bibinfo {year} {2021})}\BibitemShut {NoStop}%
\bibitem [{\citenamefont {Engdahl}\ \emph {et~al.}(2025)\citenamefont
  {Engdahl}, \citenamefont {Scammell}, \citenamefont {Efimkin},\ and\
  \citenamefont {Sushkov}}]{engdahl_theory_2025}%
  \BibitemOpen
  \bibfield  {author} {\bibinfo {author} {\bibfnamefont {J.~N.}\ \bibnamefont
  {Engdahl}}, \bibinfo {author} {\bibfnamefont {H.~D.}\ \bibnamefont
  {Scammell}}, \bibinfo {author} {\bibfnamefont {D.~K.}\ \bibnamefont
  {Efimkin}}, \ and\ \bibinfo {author} {\bibfnamefont {O.~P.}\ \bibnamefont
  {Sushkov}},\ }\href {\doibase 10.1103/PhysRevB.111.155408} {\bibfield
  {journal} {\bibinfo  {journal} {Physical Review B}\ }\textbf {\bibinfo
  {volume} {111}},\ \bibinfo {pages} {155408} (\bibinfo {year}
  {2025})}\BibitemShut {NoStop}%
\bibitem [{\citenamefont {Zou}\ \emph {et~al.}(2011)\citenamefont {Zou},
  \citenamefont {Hong},\ and\ \citenamefont {Zhu}}]{zou_effective_2011}%
  \BibitemOpen
  \bibfield  {author} {\bibinfo {author} {\bibfnamefont {K.}~\bibnamefont
  {Zou}}, \bibinfo {author} {\bibfnamefont {X.}~\bibnamefont {Hong}}, \ and\
  \bibinfo {author} {\bibfnamefont {J.}~\bibnamefont {Zhu}},\ }\href {\doibase
  10.1103/PhysRevB.84.085408} {\bibfield  {journal} {\bibinfo  {journal}
  {Physical Review B}\ }\textbf {\bibinfo {volume} {84}},\ \bibinfo {pages}
  {085408} (\bibinfo {year} {2011})}\BibitemShut {NoStop}%
\end{thebibliography}%
\end{document}